\documentclass[useAMS,usenatbib]{mn2e}

\usepackage{graphicx}        
\usepackage{natbib}
  \bibpunct{(}{)}{;}{a}{}{,} 

\newcommand{\apjs}{ApJS}
\newcommand{\apjl}{ApJL}
\newcommand{\aj}{AJ}
\newcommand{\aap}{A\&A}
\newcommand{\apj}{ApJ}
\newcommand{\mnras}{MNRAS}
\newcommand{\pasp}{PASP}
\newcommand{\nat}{Nature}

\title[The galaxies in the core of Abell 2218]{Morphologies and stellar
  populations of galaxies in the core of Abell 2218} 
\author[S.F.S\'anchez et al]{S.F.S\'anchez,
N.Cardiel$^{2,1}$,
M.A.W.Verheijen$^{3}$,
S.Pedraz$^{1,2}$ and
G.Covone$^4$\\
$^{1}$Centro Astron\'omico Hispano Alem\'an, Calar Alto, (CSIC-MPI), C/Jes\'us
 Durb\'an Rem\'on 2-2, 04004-Almer\'\i a, Spain\\
$^{2}$Departamento de Astrof\'\i sica, Facultad de F\'\i sicas, Universidad Complutense de Madrid, 28040 Madrid, Spain\\
$^{3}$Kapteyn Astronomical Institute, PO Box 800, 9700 AV Groningen, The Netherlands\\
$^4$INAF--Osservatorio Astronomico di Capodimonte, Naples, Italy\\
}
\begin{document}

\date{To be edited later (March 2006)}

\maketitle

\label{firstpage}
\begin{abstract}
  We present a study of the stellar populations and morphologies of galaxies
  in the core of the galaxy cluster Abell 2218. Integral field spectroscopy
  (IFS) observations were performed using PMAS in the PPAK mode covering a
  field-of-view of $\sim$74''$\times$64'' centred on the core of the cluster,
  in order to obtain spectroscopy of an unbiased flux limited sample 
  of cluster galaxies.  Forty-three objects were detected in the IFS data, 31
  of them with enough signal-to-noise to derive the redshift, all of them
  brighter than $I<$21.5 mag. Twenty-eight are at the redshift of the cluster
  (17 with previously unknown redshift).  Individual spectra of the cluster
  members were extracted and compared with single stellar population models to
  derive the luminosity-weighted ages and metallicities.  In addition, deep
  HST/ACS F475W, F555W, F625W and F850LP-band images centred on the cluster
  core were obtained from the HST archive ($z_{\rm lim}\sim$28 mag). A
  detailed morphological analysis of all the galaxies within the field-of-view
  of these images down to $z_{lim}<$22.5 mag was performed classifying them as
  late-type, intermediate and early-type, on the basis of their S\'ersic
  indices. The literature was scanned to look for spectroscopically confirmed
  cluster members located within the field-of-view of the ACS image.  The
  final sample of 59 galaxies comprises our reported sample of 28 galaxies in
  the core, and 31 additional galaxies in the outer regions. In addition,
  multiband broad-band photometry was extracted from the literature for all
  objects.
  
  The morphologically segregated color-magnitude diagram shows that the
  early-type galaxies cover the range of brighter and redder colors (the
  so-called ``red sequence'').  A large fraction of spiral galaxies
  ($\sim$50\%) is found, most of them fainter than the limit of previous
  studies. They cover a wide range in colors, from blue colors typical of
  Butcher-Oemler galaxies to red colors similar to those of early-type
  galaxies. This result indicates that early-type galaxies are more massive
  and have older stellar populations, while late-type galaxies are less
  massive and have a wider range of stellar populations. The distribution of
  luminosity weighted ages as a function of metallicities and luminosity
  weighted masses, and the distribution of S\'ersic indices as a function of
  the luminosity weighted masses confirm these results.  They in fact agree
  with a proposed two-step scenario for the evolution of galaxies in clusters,
  where the star formation is quenched first in the infalling spirals, after
  which a morphological transformation follows that requires larger
  time-scales. This scenario naturally explains the population faint late-type
  galaxies with old stellar populations observed in this cluster. In addition,
  an extremely blue merging galaxy system is found at the core, with the
  nominal redshift of the cluster.

\end{abstract}

\begin{keywords}
galaxies: clusters: individual: Abell 2218 - galaxies: elliptical and
lenticular, cD - galaxies: spiral - galaxies: irregular - galaxies:
fundamental parameters - galaxies: stellar content - techniques: spectroscopic 
\end{keywords}

\section{Introduction}
\label{sec:1}


Galaxy clusters have been used for decades to study the evolution of galaxies.
Being tracers of the largest density enhancements in the Universe, they are
considered the regions where galaxies formed first.  At low redshift they are
dominated by large early-type galaxies (E/S0), which colors are consistent
with a bulk formation of their stars at high redshift. Classically, it was
thought that the evolution of this dominant population through different
cosmological epochs was largely passive without signatures of important
deviations (ie., periods of more recent star formation).  However, this simple
picture has changed in the last decades.  \cite{butc83} and \cite{butc84}
showed an increase in the fraction of blue galaxies in clusters with redshift.
High-resolution ground-based and HST images have shown that they correspond to
galaxies with late-type morphologies, often disturbed or with close companions
\citep[e.g.][]{thom86,thom88,couc94,dres94,dres97}.  One observes a
morphological evolution of the galaxies in clusters, in the sense that the
fraction of E-type galaxies remains constant as a function of different
cosmological epochs, while the fraction of S0's increases at the expense of
L-types towards lower redshift \citep[e.g.][]{fasa00}.

The presence in clusters of a significant population of post-starburst
galaxies known as ``E+A'' or ``k+a'' galaxies \citep{dres83,couc87}, indicates
that star formation was active in the recent past and sharply halted at some
point during the last 1-1.5 Gyr \citep{pogg02,pogg04}. Most of these galaxies
are late type, and they show a radial distribution within the clusters that is
intermediate between the passive and the emission-line galaxy populations
\citep{pogg99}. These results lead to the hypothesis of a two-step evolution
of galaxies. A spiral galaxy is captured by the cluster and, during its
infall, suffers from gas-stripping due to the tidal forces involved
\citep[e.g.][]{marc97}. Provoked by the lack of gas, the star formation is
halted, and the galaxy evolves passively, being observed as a ``k+a''
post-starburst.  An initial short starburst process, previous to the
gas-stripping, may be also present \citep{pogg04}. After this the galaxy
evolves morphologically over larger times scales. The disk dims due to the
lack of a new generation of stars, that otherwise will dominate its
luminosity, and the galaxy transforms itself into a lenticular galaxy.  Dry
mergers between lenticular and spheroidal galaxies build up new large
spheroidals \citep[e.g.][]{bell06}

The proposed scenario suggests that a fraction of the observed S0's were
previously star-forming spirals. For these galaxies the star formation was
truncated at some time between 2 and 5 Gyr ago and it is expected that their
luminosity faded by $\sim$0.5--1.5 mag and become redder \citep{pogg01},
populating the red sequence at magnitudes fainter than the brighest 1 or 2
magnitudes \citep{pogg02}. The same process may affect dwarf galaxies captured
by the cluster, which will suffer more strongly the effects of tidal
interaction and galaxy harrasment, which induces a short but intense starburst
\citep{moor96,rako01}.  Most of the field dwarf galaxies are gas-rich rich,
blue galaxies with late-type morphologies at any redshift
\citep{cair01,bara06}. Once captured by the cluster, consumed or stripped of
its gas, it is expected that their luminosity fades and they become redder
too, populating the faint-end of the red sequence. In both cases ones expects
to find a population of red late-type galaxies, fainter than the original
early-type galaxies in the cluster. However, most of the previous studies have
found that late-type galaxies are mostly blue \citep[e.g.][]{thom86}.

To test this hypothesis, complete spectroscopic surveys of cluster galaxies
and high-resolution deep images for a proper morphological analysis of the
faintest possible targets are required.  Complete spectroscopic surveys are
difficult and time consuming, due to the requirement of deblending and placing
objects in the spectrograph multislits or masks. In particular, the blending of
different components in the core of the clusters is a problem, impossible to
handle with classical slit spectroscopy. Due to that, most of the
spectroscopic surveys are based on pre-selection of the cluster candidates,
most of them based on color or morphological selections
\cite[e.g.][]{dres99,balo99,kels00,zieg01,frit05}. Alternatively, they are
focused on galaxies in the less crowded outer regions.
These selections may affect substantially the conclusions. For these reasons
we have started an observational program aimed at obtaining Integral Field
Spectroscopy (IFS) of the core of some of the richest galaxy clusters in the
local universe, including Abell 1689 (ESO 73.A-0605) and Abell 2218. For both
of them deep HST/ACS images are available. A similar approach has been used
before by \cite{covo06}, to study the dynamical mass of low redshift galaxy
clusters.

Abell 2218 is one of the richest clusters in the Abell catalogue
\citep{abell89}, with a richness class of 4. It has a redshift of $z\sim$0.17,
and a velocity dispersion of 1370 km s$^{-1}$ \citep{kris78,borg92}. The
cluster shows a set of arc-lenses.  Detailed high-resolution X-ray maps
\citep{mcha90} and mass-concentration studies based on the properties of the
gravitational lenses have shown that the cluster has two density peaks, the
strongest of them dominated by a large cD galaxy. Several observational
programs have provided us with a large dataset of well deblended slit spectra
for the galaxies in the outer part of the cluster \citep{zieg01}, as well as
very good multiband ground-based and HST imaging \citep{borg92,smai01,zieg01}.
We focused our survey on the central arcmin region of the cluster, around the
cD galaxy.

\begin{figure}
\centering
\includegraphics[angle=270,width=8cm]{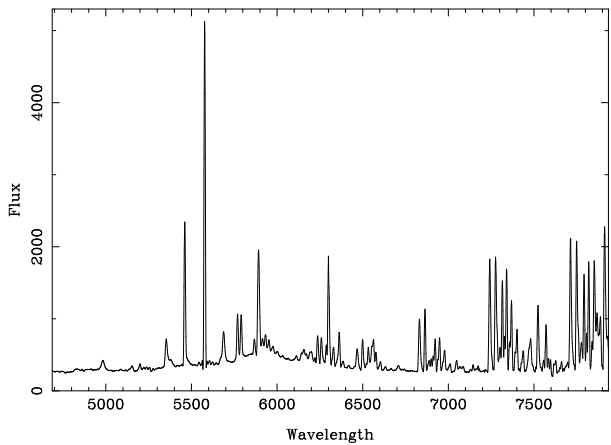}
\caption{Average night-sky emission spectrum derived by combining the spectra
  obtained through the 36 sky-fibers for the different pointings of each night.}
\label{fig:sky}       
\end{figure}

\begin{table}
 \centering
 \begin{minipage}{80mm}
  \caption{Summary of the HST/ACS images}
  \label{tab:img}
  \begin{tabular}{lccrrr}
  \hline
  Filter & Pivot $\lambda$ & Width  & Number & Exp.time & mag$_{lim}$ \\
         &  \AA          &  \AA     & Frames &   seconds    & (15$\sigma$)\\
  \hline
F475W   &4744.356 &418.80 & 3 & 4267.0 &27.42\\    
F555W   &5359.547 &375.19 & 4 & 5601.0 &28.04\\
F625W   &6310.454 &415.31 & 5 & 7013.0 &28.08\\
F775W   &7693.026 &432.03 & 8 & 10732.0 &27.46\\
F850LP  &9054.768 &538.95 &12 & 11776.0 &28.22\\
  \hline
\end{tabular}
\end{minipage}
\end{table}

The outline of this article is as follows. Section 2 describes the
observational dataset, including a description of the IFS data and the HST/ACS
images. Section 3 includes the analysis of the dataset and presents the
results. Here, Sections 3.1 and 3.2 describe the automatic morphological
analysis that we performed as opposed to classical methods. Section 3.3 shows
the method used to deblend and extract the integrated spectra of the different
detected objects from the IFS data, with the redshift determination explained
in Section 3.4. The analysis of the morphologically segregated color-magnitude
diagram is included in Section 3.5. The luminosity weighted ages and
metallicities of the galaxies, and their relation to their masses and
morphologies are discussed in Section 3.6 and 3.7. The results are discussed
in Section 4, and the conclusions are listed in Section 5.  Throughout this
article we have used a $\Lambda$-cosmology, with H$_{\rm 0}$=70 km s$^{-1}$
Mpc$^{-1}$, $\Omega_{\rm m}$=0.3 and $\Omega_{\Lambda}$=0.7.

\begin{figure*}
\centering
\includegraphics[angle=270,width=16cm]{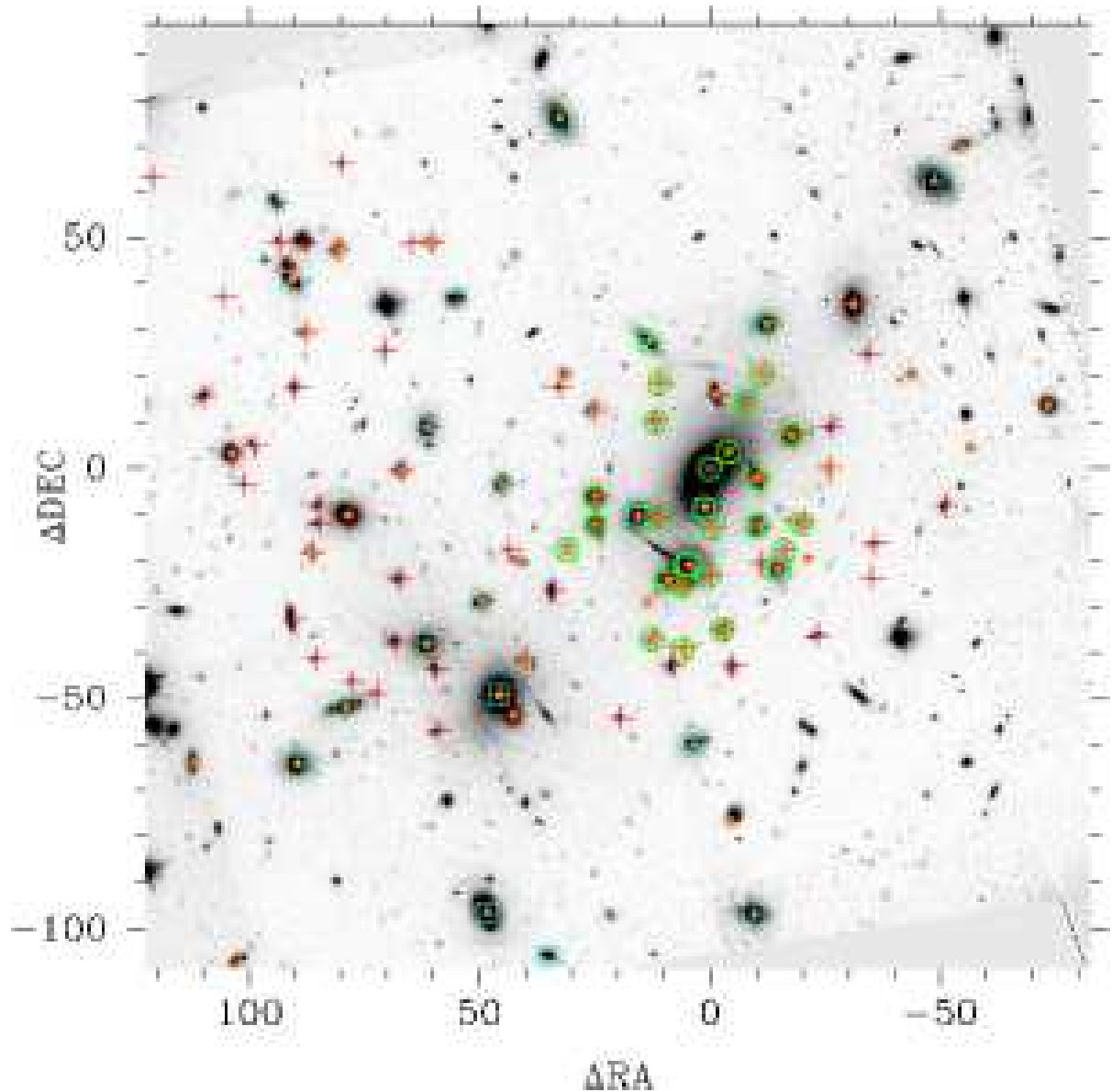}
%
%
\caption{Grayscale image of the combined F850LP-band image, together with labels
  indicating the location of the galaxies studied by Le Borgne et al. (1992)
  (orange stars), Ziegler et al. (2001) (blue squares) and Smail et al. (2001)
  (red crosses). The location of the confirmed cluster members sampled by our
  IFS data are also indicated (green circles). North is up and East left.}
\label{fig:map}       
\end{figure*}

\section{Observational Dataset}
\label{sec:2}

\subsection{IFS observations and data reduction}
\label{sec:2.1}

Observations were carried out on June the 30th and July the 6th, 2005, at the
3.5m telescope at Calar Alto Observatory with the Potsdam MultiAperture
Spectrograph, PMAS, \citep{roth05} in the PPAK mode \citep{marc04,kelz06}. The
V300 grating was used, covering a wavelength range between 4650--8000 \AA\ with
a resolution of $\sim$10 \AA\ FWHM and a sampling of 3.2 \AA/pixel. The PPAK
fiber bundle consists of 382 fibers of 2.7'' diameter each one. Of them, 331
fibers (the science fibers) are concentrated in a single hexagonal bundle
covering a field-of-view of 72''$\times$64'', with a filling factor of
$\sim$65\%. The sky is sampled by 36 additional fibers, distributed in 6
bundles of 6 fibers each one, located following a circular distribution at
$\sim$90'' from the center and at the edges of the central hexagon. The
sky-fibers are distributed between the science ones in a pseudo-slit, in order
to have a good characterization of the sky. The remaining 15 fibers are used
for calibration purposes, as described below.


For each night, three dithered exposures of 1 hour each were taken, following
a pattern of $\Delta$RA$=+$1.56'' and $\Delta$DEC$=\pm$0.78'', in order to
have a complete coverage of the science field-of-view and to increase the
spatial resolution. Each pointing was divided into shorter individual frames
for a proper cosmic-ray removal. A total of 6 hours integration time was
achieved after combining the exposures of the two nights (3 hours per night).

Data reduction was performed using R3D \citep{sanc05,sanc06}, in combination
with the IRAF package \citep{iraf}\footnote{IRAF is distributed by the
  National Optical Astronomy Observatories, which are operated by the
  Association of Universities for Research in Astronomy, Inc., under
  cooperative agreement with the National Science Foundation.} and E3D
\citep{sanc04}. The reduction consists of the standard steps for fiber-based
integral-field spectroscopy. A master bias frame was created by averaging all
the bias frames observed during the night and subtracted from the science
frames. The location of the spectra in the CCD was determined using a
continuum illuminated exposure taken before the science exposures. Each
spectrum was extracted from the science frames by coadding the flux in 5
pixels in the perpendicular direction of each fiber, and stored in a
row-staked-spectrum file RSS \citep{sanc04}. Wavelength calibration was
performed using a He lamp exposure taken at the beginning of the night, and
corrected for distortions using ThAr exposures obtained simultaneously with
the science exposures through the calibration fibers (indicated above).  The
final accuracy of the wavelength calibration was estimated by comparing the
central wavelength of the sky-emission lines with their nominal values,
finding an agreement within a range of $\sim$0.3 \AA.  Differences in the
fiber-to-fiber transmission throughput were corrected for by comparing the
wavelength-calibrated RSS science frames with the corresponding continuum
illuminated images.  Flux calibration was performed using standard calibration
stars observed throught the night (BD33+2642 and Hz44). The uncalibrated
spectra of the stars extracted from the reduced frames were then compared with
publically available calibrated spectra, determining a flux calibration
correction that was applied to the science frames, once corrected by the
differences in exposure time. The accuracy of the spectrophotometric
calibration was $\sim$0.15 mag across the wavelength range covered by our
dataset, as determined by comparing with high accurate broad-band photometry ,
as we will discuss below.  A night sky-emission spectrum was obtained for each
pointing by combining the 36 spectra obtained through the sky-fibers, and
subtracted from the 331 science-fibers spectra. Figure \ref{fig:sky} shows the
average night-sky spectrum obtained after combining all of them. Several
strong night-sky emission lines contaminate our spectra, in particular in the
reddest part ($\lambda>$7200\AA).

\begin{figure*}
\centering
\includegraphics[height=8cm]{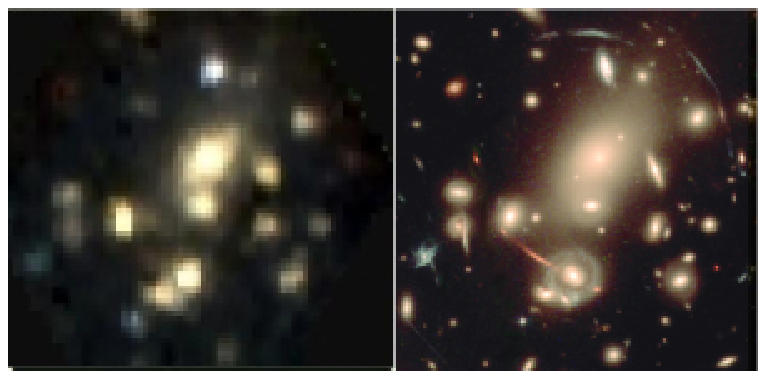}
\caption{{\it left-panel:} Three-color image created by coadding the
  flux of the final datacube through three broad-bands corresponding
  approximately to $B$, $R$ and $I$. {\it right-panel:} Similar image created
  using HST/ACS data described in the text. An electronical version of the
  full size image can be found in the webpage {\it http://www.caha.es/sanchez/abell2218/BRI\_big.jpg}}
\label{fig:1}       
\end{figure*}

The three dithered exposures were then combined, creating a single RSS frame
with 993 science spectra and the corresponding combination of the position
table.  A final datacube with 1''/pixel sampling was created for each night by
interpolating the combined RSS frame using E3D\citep{sanc04}.  This
corresponds to 2.85 kpc/pixel at the redshift of our target.  Finally, the two
datacubes obtained for each night were recentered at each wavelength,
correcting for differential atmospheric refraction, and combined using IRAF
tasks. The final datacube comprises 4828 individual spectra, covering a
field-of-view of $\sim$71''$\times$68''.



\subsection{HST/ACS images and literature data} 
\label{sub:2.2}

We obtained publically available HST/ACS images of the cluster in the F475W,
F555W, F625W, F775W and F850LP-bands from the Hubble Space Telescope archive.
Table \ref{tab:img} summarizes their main characteristics, including the
central wavelength and width of their bands \citep[obtained from][]{siri05}.
They basically correspond to images in the the classical $B$, $V$, $R$, $I$
and $z$ bands.  The images were calibrated to the Vega system using the
corresponding magnitude zeropoints and prescriptions listed in \citet{siri05}.
Table \ref{tab:img} also lists the number of individual frames acquired, each
one with a different orientation and a slightly different pointing, due to the
observing strategy of the HST/ACS camera (in order to cover the gap in-between
the two different CCDs). The acquired frames were already processed by the
HST/ACS reduction pipeline, which includes bias and dark current correction,
flat-fielding, quantum efficiency correction, field-distorsion correction, the
derivation of the astrometric-solution and flux calibration.  We correct each
individual from for cosmic rays, realigned to the standard North-East and
finally combined into a single image per filter after a sky-background
determination and subtraction. The final exposure time and limiting magnitude
of each image, estimated by looking for the faintest object detected at
$\sim$15$\sigma$ over the background, are listed in Table \ref{tab:img}. All
the images are much deeper than our IFS data, as we will see below in detail.
In particular this is the case for the F850LP-band image.  This image is much
deeper than the rest, being as deep as the corresponding image of UDF
\citep[e.g.][]{coe06}, and therefore one of the deepest images ever taken of
this cluster. The cluster core is centred on one of the ACS chips for this
particular image, and the field-of-view of our IFS data is completely covered
by it. This high resolution ACS F850LP-band image will be used for a detailed
morphological analysis.

We also collected broad-band photometry and redshift information of the
cluster galaxies available in the literature. \citet{borg92} performed a
photometric and spectroscopic survey of the galaxies in a field of
4$\arcmin\times$4$\arcmin$ centred on Abell 2218. This survey comprises
broad-band photometry in 5 bands ($B$,$g$,$r$,$i$ and $z$ filters) for 729
objects, and redshift determination for 66 of them, 51 cluster members.  They
calibrated their photometric data using two different systems, the Johnson one
for the $B$-band \citep{john53}, and the Thuan-Gunn for the $g$,$r$,$i$ and
$z$-band \citep{thua76}.  We did not perfom any recalibration of their data to
a common photometric system.  These data were initially used to compare with
the corresponding intensity of the individual spectra of the objects detected
in our IFU data, and for doing so we just needed to use the correspoding
Johnson and Thuan-Gunn effective flux \cite[e.g][]{fuku95}. The accuracy of
their photometry ranges from 0.1 mag to 0.5 mag from the brighter ($B\sim$20
mag) to the fainter ($B\sim$ 26 mag) objects. The spectroscopic data lacks
the required quality needed to perform an accurate analysis of the
age-metallicity indicators, like the equivalent widths of H$\beta$ and Mgb, but
they provide high quality redshift determinations.

\citet{zieg01} reported on multislit moderate-resolution spectroscopic
observations of 48 color-selected early-type members of the cluster. They also
present high-quality multiband photometry in 4 bands ($U$, $B$, $V$ and $I$),
derived from groud-based observations, and a morphological analysis of 19 of
the 48 galaxies based on HST/WFPC2 observations (F702W-band, $\sim R$-band).
Finally, \cite{smai01} present a photometric study of the ages and
metallicities of 81 early-type cluster candidates (33 with confirmed
redshifts), based on ground-based $K$-band and HST/WFPC multiband images, in
the F450W, F606W and F814W-bands. In both cases, their photometric data were
calibrated in the Vega-based system. Many other studies have targeted this
cluster since its redshift confirmation by \cite{kris78}, and the early
photometric studies by \cite{butc83}. However, in this paper we only use the
photometric and spectroscopic data of the articles listed above.

Figure \ref{fig:map} shows a greyscale of the combined HST/ACS F850LP-band
image, together with labels indicating the location of the galaxies studied by
Le Borgne et al. (1992) (orange stars), Ziegler et al. (2001) (blue squares)
and Smail et al. (2001) (red crosses). The location of the confirmed cluster
members sampled by our IFS data are also indicated (green circles). As we will
see below, most of them have not been previously targeted by any spectroscopic
survey.

\begin{table*}
 \centering
 \begin{minipage}{160mm}
  \caption{Summary of the photometric analysis.}
  \label{tab:phot}
  \begin{tabular}{rrrccccc}
  \hline
  ID & RA & DEC & $B_{475}$ & $V_{555}$ & $R_{625}$ & $I_{775}$ & $z_{850}$ \\
  \hline
  1 &248.9537297 &+66.2117906 &18.09$\pm$0.02 &17.29$\pm$0.02 &16.52$\pm$0.01 &15.85$\pm$0.01 &15.44$\pm$0.01\\
247 &248.9853560 &+66.1981066 &18.73$\pm$0.03 &17.94$\pm$0.02 &17.14$\pm$0.02 &16.44$\pm$0.01 &16.08$\pm$0.02\\
169 &248.9870850 &+66.1850276 &18.86$\pm$0.03 &18.09$\pm$0.03 &17.34$\pm$0.02 &16.70$\pm$0.02 &16.19$\pm$0.02\\
512 &248.9569254 &+66.2059389 &18.92$\pm$0.03 &18.32$\pm$0.03 &17.51$\pm$0.02 &16.80$\pm$0.02 &16.44$\pm$0.02\\
554 &248.9546845 &+66.2094133 &19.53$\pm$0.04 &18.87$\pm$0.04 &18.26$\pm$0.03 &17.56$\pm$0.03 &17.00$\pm$0.03\\
172 &249.0078962 &+66.2089431 &19.22$\pm$0.04 &18.46$\pm$0.03 &17.67$\pm$0.02 &16.96$\pm$0.02 &16.47$\pm$0.02\\
501 &248.9646342 &+66.2088796 &19.40$\pm$0.04 &18.67$\pm$0.04 &17.93$\pm$0.03 &17.12$\pm$0.02 &16.61$\pm$0.03\\
561 &248.9762849 &+66.2328158 &19.33$\pm$0.04 &18.62$\pm$0.04 &17.92$\pm$0.02 &17.17$\pm$0.02 &16.65$\pm$0.03\\
189 &248.9961745 &+66.2011969 &19.57$\pm$0.05 &18.75$\pm$0.04 &17.93$\pm$0.02 &17.27$\pm$0.02 &16.78$\pm$0.03\\
103 &249.0154109 &+66.1939177 &19.50$\pm$0.05 &18.72$\pm$0.04 &17.91$\pm$0.02 &17.25$\pm$0.02 &16.68$\pm$0.03\\
132 &249.0082842 &+66.1973755 &19.57$\pm$0.05 &18.76$\pm$0.04 &17.95$\pm$0.03 &17.27$\pm$0.02 &16.73$\pm$0.03\\
555 &248.9468602 &+66.2111799 &19.92$\pm$0.05 &19.19$\pm$0.05 &18.44$\pm$0.03 &17.69$\pm$0.03 &17.14$\pm$0.03\\
201 &249.0162635 &+66.2230941 &19.61$\pm$0.05 &18.83$\pm$0.04 &18.09$\pm$0.03 &17.39$\pm$0.02 &16.89$\pm$0.03\\
  2 &248.9453711 &+66.2204800 &19.89$\pm$0.06 &19.10$\pm$0.05 &18.35$\pm$0.03 &17.65$\pm$0.03 &17.12$\pm$0.03\\
256 &248.9833226 &+66.1968621 &19.89$\pm$0.05 &19.20$\pm$0.05 &18.46$\pm$0.03 &17.80$\pm$0.03 &17.20$\pm$0.04\\
509 &248.9602122 &+66.2050146 &20.20$\pm$0.06 &19.44$\pm$0.05 &18.43$\pm$0.03 &17.73$\pm$0.03 &17.26$\pm$0.04\\
349 &248.9959951 &+66.2142513 &20.00$\pm$0.06 &19.23$\pm$0.05 &18.54$\pm$0.03 &17.77$\pm$0.03 &17.27$\pm$0.04\\
126 &249.0253609 &+66.2126854 &19.90$\pm$0.06 &19.12$\pm$0.05 &18.41$\pm$0.03 &17.73$\pm$0.03 &17.26$\pm$0.04\\
570 &248.9630196 &+66.2194216 &20.11$\pm$0.06 &19.35$\pm$0.05 &18.62$\pm$0.04 &17.93$\pm$0.03 &17.41$\pm$0.04\\
523 &248.9871643 &+66.2387374 &$---$ &$---$ &$---$ &$---$ &17.39$\pm$0.04\\
604 &248.9528976 &+66.2162360 &19.41$\pm$0.04 &18.95$\pm$0.04 &18.35$\pm$0.03 &17.94$\pm$0.03 &17.49$\pm$0.04\\
599 &248.9437771 &+66.2056763 &20.23$\pm$0.07 &19.40$\pm$0.05 &18.65$\pm$0.04 &18.00$\pm$0.03 &17.46$\pm$0.04\\
658 &248.9415886 &+66.2137478 &20.29$\pm$0.07 &19.53$\pm$0.06 &18.80$\pm$0.04 &18.11$\pm$0.04 &17.59$\pm$0.04\\
337 &248.9878980 &+66.2037120 &20.53$\pm$0.08 &19.76$\pm$0.06 &18.97$\pm$0.04 &18.35$\pm$0.04 &17.64$\pm$0.05\\
596 &248.9468067 &+66.2083220 &20.68$\pm$0.08 &19.83$\pm$0.07 &19.07$\pm$0.05 &18.37$\pm$0.04 &17.77$\pm$0.05\\
462 &248.9709853 &+66.2100529 &20.30$\pm$0.07 &19.49$\pm$0.06 &18.85$\pm$0.04 &18.21$\pm$0.04 &17.86$\pm$0.05\\
471 &248.9564109 &+66.1951060 &20.64$\pm$0.08 &19.83$\pm$0.07 &19.03$\pm$0.04 &18.46$\pm$0.04 &17.86$\pm$0.05\\
302 &248.9999910 &+66.2116528 &20.68$\pm$0.08 &19.91$\pm$0.07 &19.18$\pm$0.05 &18.53$\pm$0.04 &17.79$\pm$0.05\\
388 &248.9849322 &+66.2108143 &20.41$\pm$0.07 &19.74$\pm$0.06 &19.08$\pm$0.05 &18.48$\pm$0.04 &17.97$\pm$0.05\\
449 &248.9708088 &+66.2082924 &20.62$\pm$0.56 &19.85$\pm$0.10 &19.05$\pm$0.04 &18.39$\pm$0.04 &18.01$\pm$0.05\\
 72 &249.0309189 &+66.1939739 &$---$ &$---$ &$---$ &$---$ &17.89$\pm$0.05\\
408 &248.9916711 &+66.2220150 &20.62$\pm$0.08 &19.91$\pm$0.07 &19.19$\pm$0.05 &$---$ &17.95$\pm$0.05\\
 95 &249.0389514 &+66.2155899 &20.96$\pm$0.09 &20.24$\pm$0.08 &19.53$\pm$0.06 &18.87$\pm$0.05 &18.33$\pm$0.06\\
798 &249.0240772 &+66.1820752 &$---$ &$---$ &$---$ &$---$ &18.41$\pm$0.07\\
511 &248.9579508 &+66.2047618 &20.35$\pm$0.11 &19.72$\pm$0.10 &19.25$\pm$0.06 &18.68$\pm$0.06 &18.51$\pm$0.07\\
493 &248.9754372 &+66.2174786 &21.44$\pm$0.12 &20.74$\pm$0.10 &19.99$\pm$0.07 &19.33$\pm$0.07 &18.64$\pm$0.07\\
643 &248.9401830 &+66.2084529 &21.45$\pm$0.12 &20.67$\pm$0.10 &19.97$\pm$0.07 &19.29$\pm$0.06 &18.59$\pm$0.07\\
329 &249.0093949 &+66.2250084 &21.19$\pm$0.11 &20.49$\pm$0.09 &19.69$\pm$0.06 &19.00$\pm$0.06 &18.52$\pm$0.07\\
549 &248.9519752 &+66.2020070 &21.33$\pm$0.11 &20.65$\pm$0.10 &19.94$\pm$0.07 &19.23$\pm$0.06 &18.76$\pm$0.08\\
632 &248.9484012 &+66.2157872 &21.47$\pm$0.14 &20.50$\pm$0.10 &19.84$\pm$0.06 &19.21$\pm$0.06 &18.89$\pm$0.07\\
357 &248.9817003 &+66.2002660 &21.85$\pm$0.14 &20.54$\pm$0.09 &19.73$\pm$0.06 &19.05$\pm$0.06 &18.53$\pm$0.07\\
560 &248.9510744 &+66.2127998 &22.45$\pm$0.19 &21.81$\pm$0.17 &21.06$\pm$0.12 &20.49$\pm$0.12 &20.00$\pm$0.10\\
196 &249.0143811 &+66.2200223 &21.71$\pm$0.13 &20.99$\pm$0.12 &20.26$\pm$0.08 &19.57$\pm$0.07 &19.01$\pm$0.09\\
384 &248.9824979 &+66.2061905 &21.46$\pm$0.12 &20.83$\pm$0.11 &20.17$\pm$0.08 &19.57$\pm$0.07 &19.09$\pm$0.09\\
411 &248.9954237 &+66.2254124 &21.75$\pm$0.14 &20.99$\pm$0.12 &20.27$\pm$0.08 &19.56$\pm$0.07 &18.91$\pm$0.08\\
603 &248.9619939 &+66.2146043 &21.83$\pm$0.14 &21.10$\pm$0.12 &20.37$\pm$0.09 &19.73$\pm$0.08 &19.13$\pm$0.09\\
533 &248.9615077 &+66.2088133 &22.07$\pm$0.15 &21.39$\pm$0.14 &20.60$\pm$0.10 &19.78$\pm$0.08 &19.30$\pm$0.09\\
619 &248.9427500 &+66.2068096 &21.94$\pm$0.15 &21.21$\pm$0.13 &20.58$\pm$0.09 &19.93$\pm$0.09 &19.30$\pm$0.10\\
484 &248.9624573 &+66.2014471 &21.84$\pm$0.14 &21.24$\pm$0.13 &20.52$\pm$0.09 &19.97$\pm$0.09 &19.50$\pm$0.10\\
153 &249.0133589 &+66.2067467 &21.83$\pm$0.14 &21.16$\pm$0.13 &20.41$\pm$0.09 &19.79$\pm$0.08 &19.24$\pm$0.10\\
513 &248.9540625 &+66.2052672 &22.31$\pm$0.23 &21.58$\pm$0.15 &20.64$\pm$0.18 &19.64$\pm$0.18 &19.04$\pm$0.09\\
507 &248.9578870 &+66.2008693 &22.40$\pm$0.20 &21.68$\pm$0.18 &21.03$\pm$0.12 &20.39$\pm$0.11 &19.84$\pm$0.13\\
431 &248.9752130 &+66.2068778 &20.89$\pm$0.09 &20.60$\pm$0.10 &20.27$\pm$0.08 &19.97$\pm$0.09 &19.78$\pm$0.13\\
659 &248.9460977 &+66.2174923 &23.69$\pm$0.34 &22.91$\pm$0.29 &22.14$\pm$0.20 &21.46$\pm$0.18 &20.81$\pm$0.17\\
575 &248.9613337 &+66.2169292 &22.94$\pm$0.24 &22.05$\pm$0.20 &21.43$\pm$0.15 &20.79$\pm$0.13 &20.36$\pm$0.17\\
574 &248.9534579 &+66.2081824 &23.14$\pm$0.32 &22.48$\pm$0.28 &21.74$\pm$0.16 &21.02$\pm$0.14 &20.47$\pm$0.16\\
  \hline
\end{tabular}
\end{minipage}
\end{table*}

\subsection{The depth of our IFS data}
\label{sub:2.3}

To obtain a rough estimate of the depth of our IFS data, we compared them with
available deep optical images. Figure \ref{fig:1} shows a three-color image
created by coadding the flux of the final datacube through three broad-bands
corresponding approximately to $B$, $R$ and $I$ (left panel), together with a
similar image created using the previously descrived HST/ACS dataset.  It is
interesting to note the similarities between the two images, despite the fact
that our datacube has a final FWHM for point-like sources of $\sim$2.5'', much
larger than the one from the ACS. Based on the $I$-band magnitude of the
faintest detected object, we roughly estimate the detection limit for our IFS
data I$_{lim}\sim$22.4 mag. However, we only get high enough quality spectra
for the proposed study for objects brighter than I$<$21.5 mag (signal-to-noise
ratio $>$5$\sigma$ per spectral pixel), as we will see below. This limit is
far beyond previous spectroscopic studies on this cluster, around
I$_{lim}\sim$18.7 mag \citep{zieg01}.

\section{Analysis and Results}

\subsection{Photometric analysis}

We used the previously described HST/ACS images to obtain homogeneus multiband
photometry for all the detected object in field of Abell 2218. This homogeneus
photometric dataset is fundamental to estimate the accuracy of the the
spectrophotometric calibration of our IFS data and to study the stellar
populations of the galaxies in the cluster. We use {\tt SExtractor}
\citep{bert96} to detect objects in the images up to 15$\sigma$ over the sky
(20 connected pixels with a signal-to-noise of at least 2$\sigma$ each one).
This limit is very relaxed, and most probably we are losing many real faint
objects by using it. However, we are not interested in the faintest targets,
most probably high-z objects or arclenses (see Fig 1), but only in the cluster
members. For each detected target we obtained aperture photometry in each
band, using the MAG\_BEST parameter derived by SExtractor. The little
variations of the HST/ACS PSF compared to the size of our targets of interest
guarantees homogeneous photometric dataset. Table \ref{tab:phot} lists the
results from this analysis for the 59 galaxies confirmed to be cluster
members, using the published data listed above or the results from our
spectroscopic survey (see below).  For each target it includes the SExtractor
ID, the coordinates (RA and DEC), and the photometry in each band plus errors.
Due to the change of orientation and slightly different pointing of the
individual HST/ACS frames, the F850LP-band covers a slightly larger area than
the images in the other bands, and some of the objects were only detected in
this band.

\begin{figure}
\centering
\includegraphics[angle=270,width=8cm]{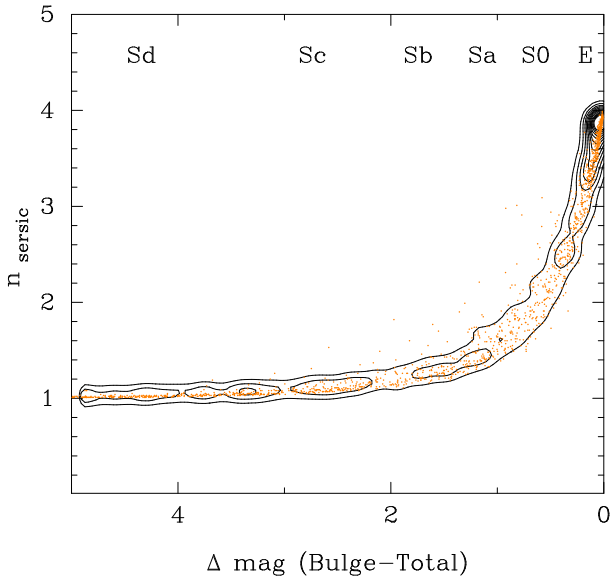}
%
%
\caption{Results from the simulations. Solid-dots show the recovered 
  S\'ersic indices of the simulated galaxies as a function of the magnitude
  difference between the bulge and total components ($\Delta$mag). Overplotted
  countours show the density distribution of the plotted points.}
\label{fig:sim}       
\end{figure}

\subsection{Morphological analysis}
\label{sec:morph}

One of the goals of the present study is to determine the stellar populations
of galaxies in the cluster as a function of their morphologies, in order to
understand their origin and evolution. To achive this goal we performed a
morphological analysis of the HST/ACS F850LP images, the deepest of our
dataset, following the prescriptions presented in \citet{coe06}, where similar
images were analyzed.  This analysis recovers, for each galaxy, the S\'ersic
index, the integrated magnitude, the effective radius, the ellipticity, the
position angle, and the asymmetry index.

For undisturbed galaxies the S\'ersic index $n$ correlates fairly well with
morphological type \citep[e.g.][]{andr95,bell04}, in the sense that galaxies
with high $n$ have higher central concentrations (i.e., bulge dominated
galaxies) than galaxies with low $n$ (i.e., disk dominated galaxies). Less
well behaved galaxies, including mergers and irregulars, generally do not have
well defined S\'ersic indices. Fortunatelly these galaxies can generally be
weeded out (or selected for) by measuring their large asymmetries
\citep[e.g.][]{cons03}. 
 
In summary the morphological analysis consists of the following steps: First
we use {\tt SExtractor} \citep{bert96} to detect galaxies in the combined
F850LP-band image.  For each detected galaxy we recover the physical and image
coordinates (RA,DEC, X and Y), the integrated magnitude (mag\_best), the
scale-length ($r_{50}$), the position angle and the ellipticity. Then, we
create a postage stamp 5$\times r_{50}$ on a side, for the brighest galaxy.
Within that postage stamp, neighbouring galaxies are masked out using
ellipses, each ellipse given a minor axis length $b = 2\times r_{50}$ for that
galaxy. Using {\tt galfit} \citep{peng02}, the brightest galaxy is fitted to a
single component S\'ersic model $\Sigma (R) = \Sigma_e\times \exp(-\kappa_n
[(R/R_e)^{1/n}-1])$, where $R_e$ is the effective or half-light radius,
$\Sigma_e$ is the effective surface density, $\Sigma\left(R\right)$ is the
surface density as a function of radius and $\kappa=\kappa\left(n\right)$ is a
normalization constant. The fit is constrained to 0.2$<n<$8 and 0.3$<R_e<$500
pixels, and the centroid is confined to within 2 pixels of the position
derived by {\tt SExtractor}. As initial guesses for the {\tt galfit}
parameters, we use the {\tt SExtractor} output parameters. Lacking estimates
for the S\'ersic index from {\tt SExtractor}, we start all fits with $n=1.5$.
{\tt galfit} convolves the model with the required PSF. Prior to the fit, we
created a PSF image by combining postage stamps of unsaturated point-like
objects extracted from the combined F850LP-band image.

Having been calculated for the brightest galaxy, the S\'ersic model is
substracted from the F850LP-band image. This subtraction benefits the
subsequent modelling of fainter nearby galaxies. We then proceed to model the
second brightest galaxy, and continue in order of decreasing brightness,
modelling and subtracting every galaxy in the {\tt SExtractor} catalogue. {\tt
  galfit} is able of simultaneous fit several targets, producing reasonable
results, in most of the cases. However, in this case the fitting process is
dominated by the brighter targets, and flux of the fainter ones is more poorly
recovered. We performed extensive tests before adopting the described
procedure. In addition, we determine the asymmetry index (A), following the
prescriptions in \citet{coe06}. Of the 802 objects detected by {\tt
  SExtractor} in the F850LP-band image brighter than $z<$26.5 mag, we applied
{\tt galfit} to the 241 galaxies brighter than $z<$22.5 mags. The error in the
derived Sersic index is restricted to a range of $\pm$0.25, based on the
simulations presented in \citet{coe06}, and therefore a morphological
segregation based on that parameter is reliable. The remaining galaxies are
far beyond the depth of our spectroscopic data, and therefore excluded. Table
\ref{tab:morph} lists the resulting fitted parameters and their uncertainties
for the 59 galaxies confirmed to be cluster members, using the published data
listed above or the results from our spectroscopic survey (see below). It
includes, for each galaxy, the {\tt SExtractor} identification, the
$\chi^2/\nu$ of the fitting procedure, the $z$-band magnitude, the effective
radius (R$_{\rm e}$), the axis ratio ($a/b$), the position angle ($\theta$),
the S\'ersic index ($n_s$) and the Asymmetry index (A).  The listed errors
were estimated based on the results of the simulations presented in
\citet{coe06}. The error in the Asymmetry index is dificult to quantify for
individual objects, however, based on the quoted simulations, we estimate it
to be lower than $\sim$0.1 for the galaxies of our sample.  Of all these
galaxies, only one has a strong Asymmetry index ($\sim$0.8).  This galaxy is
covered by the field-of-view of our IFS data, and it will be discussed later
in detail.

\begin{table*}
 \centering
 \begin{minipage}{160mm}
  \caption{Summary of the morphological analysis.}
  \label{tab:morph}
  \begin{tabular}{rrrrrrrr}
  \hline
  ID & $\chi^2/\nu$ & 
        z$_{850}$ & R$_{\rm e}$ ($\arcsec$) & a/b & $\theta$ & $n_s$ & A \\
  \hline
  1 &0.017 &16.10$\pm$0.02 &12.06$\pm$0.05 &0.49$\pm$0.01 & 37$\pm$ 4  &1.63$\pm$0.06 &0.093 \\ 
247 &0.023 &16.21$\pm$0.02 & 5.18$\pm$0.05 &0.85$\pm$0.01 & 51$\pm$ 4  &1.88$\pm$0.06 &0.175 \\ 
169 &0.024 &17.07$\pm$0.02 & 1.66$\pm$0.06 &0.49$\pm$0.02 & 89$\pm$ 4  &1.48$\pm$0.07 &0.043 \\ 
512 &0.033 &17.20$\pm$0.02 & 1.51$\pm$0.06 &0.77$\pm$0.02 &-37$\pm$ 8  &2.03$\pm$0.07 &0.156 \\ 
554 &0.034 &17.37$\pm$0.02 & 0.63$\pm$0.06 &0.80$\pm$0.02 &-10$\pm$ 5  &1.80$\pm$0.07 &0.321 \\ 
172 &0.016 &16.84$\pm$0.02 & 1.64$\pm$0.06 &0.82$\pm$0.01 &-46$\pm$ 9  &2.95$\pm$0.06 &0.056 \\ 
501 &0.026 &15.92$\pm$0.02 & 0.81$\pm$0.05 &0.84$\pm$0.01 & 50$\pm$ 5  &2.86$\pm$0.05 &0.045 \\ 
561 &0.020 &17.25$\pm$0.02 & 1.52$\pm$0.06 &0.69$\pm$0.02 &-81$\pm$12  &1.47$\pm$0.07 &0.081 \\ 
189 &0.019 &17.14$\pm$0.02 & 2.69$\pm$0.06 &0.90$\pm$0.02 &-59$\pm$10  &3.93$\pm$0.07 &0.088 \\ 
103 &0.011 &16.93$\pm$0.02 & 1.17$\pm$0.05 &0.94$\pm$0.01 &-26$\pm$ 5  &2.69$\pm$0.06 &0.042 \\ 
132 &0.018 &17.28$\pm$0.02 & 1.20$\pm$0.06 &0.46$\pm$0.02 & -6$\pm$ 5  &2.78$\pm$0.07 &0.058 \\ 
201 &0.027 &17.49$\pm$0.02 & 0.89$\pm$0.06 &0.50$\pm$0.02 &-61$\pm$10  &1.91$\pm$0.07 &0.132 \\ 
  2 &0.013 &17.43$\pm$0.02 & 0.96$\pm$0.06 &0.80$\pm$0.02 &-16$\pm$ 5  &2.63$\pm$0.07 &0.050 \\ 
256 &0.022 &17.64$\pm$0.02 & 0.51$\pm$0.06 &0.85$\pm$0.02 & 75$\pm$ 5  &1.74$\pm$0.07 &0.078 \\ 
509 &0.028 &17.46$\pm$0.02 & 1.09$\pm$0.06 &0.66$\pm$0.02 &-33$\pm$ 7  &3.19$\pm$0.07 &0.129 \\ 
349 &0.016 &17.98$\pm$0.02 & 0.86$\pm$0.07 &0.70$\pm$0.02 & 71$\pm$ 4  &2.23$\pm$0.07 &0.077 \\ 
126 &0.033 &17.44$\pm$0.02 & 0.68$\pm$0.06 &0.95$\pm$0.02 &-83$\pm$12  &2.00$\pm$0.07 &0.159 \\ 
570 &0.020 &18.32$\pm$0.03 & 1.15$\pm$0.08 &0.37$\pm$0.03 &-63$\pm$10  &1.81$\pm$0.08 &0.054 \\ 
523 &0.018 &17.78$\pm$0.02 & 0.70$\pm$0.07 &0.57$\pm$0.02 &-88$\pm$13  &1.74$\pm$0.07 &0.047 \\ 
 73 &0.018 &17.90$\pm$0.02 & 0.83$\pm$0.07 &0.81$\pm$0.02 & 38$\pm$ 5  &2.02$\pm$0.07 &0.077 \\ 
604 &0.021 &18.09$\pm$0.03 & 1.13$\pm$0.08 &0.44$\pm$0.03 & 84$\pm$ 4  &0.72$\pm$0.08 &0.129 \\ 
599 &0.019 &17.76$\pm$0.02 & 1.04$\pm$0.07 &0.91$\pm$0.02 & 20$\pm$ 5  &3.94$\pm$0.07 &0.051 \\ 
658 &0.016 &17.87$\pm$0.02 & 1.01$\pm$0.07 &0.88$\pm$0.02 & 51$\pm$ 6  &2.69$\pm$0.07 &0.081 \\ 
337 &0.015 &17.96$\pm$0.02 & 0.59$\pm$0.08 &0.78$\pm$0.02 &  5$\pm$ 4  &2.49$\pm$0.07 &0.050 \\ 
596 &0.015 &18.96$\pm$0.03 & 0.65$\pm$0.09 &0.64$\pm$0.05 & 58$\pm$ 6  &2.13$\pm$0.08 &0.067 \\ 
210 &0.041 &18.19$\pm$0.03 & 0.47$\pm$0.08 &0.59$\pm$0.03 &-21$\pm$ 6  &2.03$\pm$0.08 &0.823 \\ 
462 &0.021 &18.21$\pm$0.03 & 1.35$\pm$0.08 &0.84$\pm$0.03 &-21$\pm$ 7  &2.35$\pm$0.08 &0.102 \\ 
471 &0.015 &18.12$\pm$0.03 & 0.97$\pm$0.08 &0.89$\pm$0.03 &  3$\pm$ 5  &2.11$\pm$0.08 &0.118 \\ 
302 &0.014 &18.14$\pm$0.03 & 0.48$\pm$0.08 &0.93$\pm$0.03 &  5$\pm$ 6  &2.01$\pm$0.08 &0.048 \\ 
388 &0.015 &18.25$\pm$0.03 & 0.71$\pm$0.08 &0.76$\pm$0.03 & 45$\pm$ 5  &1.68$\pm$0.08 &0.065 \\ 
449 &0.021 &18.57$\pm$0.03 & 1.60$\pm$0.08 &0.87$\pm$0.04 & 89$\pm$ 6  &1.03$\pm$0.08 &0.316 \\ 
 72 &0.019 &18.64$\pm$0.03 & 0.73$\pm$0.09 &0.44$\pm$0.04 & 72$\pm$ 5  &2.13$\pm$0.08 &0.057 \\ 
408 &0.017 &18.59$\pm$0.03 & 0.72$\pm$0.08 &0.74$\pm$0.04 &-12$\pm$ 6  &1.46$\pm$0.08 &0.152 \\ 
 95 &0.015 &18.41$\pm$0.03 & 0.63$\pm$0.08 &0.85$\pm$0.03 & 80$\pm$ 6  &1.45$\pm$0.08 &0.062 \\ 
798 &0.020 &19.02$\pm$0.04 & 0.88$\pm$0.10 &0.45$\pm$0.06 &  9$\pm$ 5  &0.81$\pm$0.09 &0.068 \\ 
511 &0.029 &18.74$\pm$0.03 & 0.57$\pm$0.09 &0.53$\pm$0.04 &-67$\pm$12  &2.35$\pm$0.08 &0.169 \\ 
493 &0.018 &18.99$\pm$0.03 & 0.45$\pm$0.09 &0.72$\pm$0.06 &-42$\pm$ 9  &1.56$\pm$0.08 &0.058 \\ 
643 &0.014 &19.09$\pm$0.04 & 0.45$\pm$0.10 &0.99$\pm$0.06 & 84$\pm$13  &1.74$\pm$0.09 &0.094 \\ 
329 &0.015 &18.71$\pm$0.03 & 0.76$\pm$0.09 &0.77$\pm$0.04 & 41$\pm$ 6  &1.46$\pm$0.08 &0.071 \\ 
549 &0.012 &19.04$\pm$0.04 & 0.63$\pm$0.10 &0.95$\pm$0.06 &-35$\pm$ 9  &1.77$\pm$0.09 &0.067 \\ 
632 &0.017 &19.03$\pm$0.04 & 0.37$\pm$0.10 &0.73$\pm$0.06 &-82$\pm$13  &1.49$\pm$0.09 &0.102 \\ 
357 &0.018 &18.70$\pm$0.03 & 0.60$\pm$0.09 &0.67$\pm$0.04 & -9$\pm$ 6  &1.07$\pm$0.08 &0.184 \\ 
560 &0.049 &18.39$\pm$0.03 & 0.17$\pm$0.08 &0.87$\pm$0.03 &-74$\pm$12  &0.53$\pm$0.08 &0.469 \\ 
503 &0.013 &19.33$\pm$0.04 & 0.83$\pm$0.10 &0.80$\pm$0.07 & 13$\pm$ 6  &0.97$\pm$0.09 &0.133 \\ 
196 &0.016 &19.14$\pm$0.04 & 0.51$\pm$0.10 &0.67$\pm$0.06 & -8$\pm$ 6  &1.38$\pm$0.09 &0.079 \\ 
384 &0.017 &19.53$\pm$0.06 & 0.77$\pm$0.12 &0.43$\pm$0.08 &-45$\pm$10  &1.29$\pm$0.09 &0.067 \\ 
411 &0.010 &18.86$\pm$0.03 & 0.39$\pm$0.09 &0.96$\pm$0.04 &-81$\pm$13  &1.88$\pm$0.08 &0.080 \\ 
603 &0.012 &19.59$\pm$0.06 & 0.49$\pm$0.12 &0.96$\pm$0.08 & 43$\pm$ 9  &1.19$\pm$0.09 &0.059 \\ 
533 &0.015 &18.87$\pm$0.03 & 0.27$\pm$0.09 &0.94$\pm$0.04 & 65$\pm$ 9  &1.44$\pm$0.08 &0.045 \\ 
619 &0.015 &19.55$\pm$0.06 & 0.41$\pm$0.12 &0.74$\pm$0.08 &  5$\pm$ 6  &1.29$\pm$0.09 &0.069 \\ 
484 &0.017 &19.71$\pm$0.06 & 0.75$\pm$0.13 &0.51$\pm$0.08 & 36$\pm$ 5  &1.18$\pm$0.09 &0.078 \\ 
153 &0.013 &19.69$\pm$0.06 & 0.69$\pm$0.13 &0.72$\pm$0.08 &-76$\pm$13  &1.34$\pm$0.09 &0.062 \\ 
513 &0.024 &20.04$\pm$0.08 & 1.15$\pm$0.17 &0.50$\pm$0.10 & 39$\pm$10  &2.31$\pm$0.10 &0.194 \\ 
507 &0.013 &20.27$\pm$0.09 & 0.38$\pm$0.18 &0.85$\pm$0.11 & 74$\pm$10  &1.31$\pm$0.10 &0.061 \\ 
431 &0.014 &20.20$\pm$0.09 & 1.10$\pm$0.18 &0.85$\pm$0.11 & 32$\pm$11  &0.89$\pm$0.10 &0.218 \\ 
803 &0.017 &21.21$\pm$0.16 & 0.67$\pm$0.26 &0.86$\pm$0.15 & 64$\pm$17  &0.40$\pm$0.15 &0.240 \\ 
659 &0.019 &20.66$\pm$0.11 & 0.25$\pm$0.22 &0.81$\pm$0.12 &-83$\pm$17  &1.68$\pm$0.12 &0.085 \\ 
575 &0.010 &20.86$\pm$0.09 & 0.75$\pm$0.24 &0.85$\pm$0.12 &-40$\pm$13  &1.39$\pm$0.13 &0.119 \\ 
574 &0.011 &20.49$\pm$0.06 & 0.27$\pm$0.22 &0.84$\pm$0.11 &-89$\pm$18  &1.15$\pm$0.12 &0.073 \\ 
  \hline
\end{tabular}
\end{minipage}
\end{table*}

\subsection{The S\'ersic indices and the Hubble sequence}
\label{sec:hubb}

Our morphological classification scheme is based on the S\'ersic index of the
galaxy profile. We consider this automatic classification scheme more straight
forward, easy to reproduce, and less subjetive than the classical {\it by-eye}
classification \citep[read][ for an illustration of the problem]{wolf05}.
However, most of the previous studies on the morphologies of galaxies in
clusters are based on {\it by-eye} classifications following the Hubble scheme
\citep{hubb26}, revised by \cite{vauc76}. Normally, they subdivide the
galaxies in their different Hubble types, or at least in three groups: pure
spheroidals (E-type), lenticulars (S0), and spirals (L-type)
\citep[e.g.][]{fasa00,pogg01,zieg01,smai01}. In order to compare with previous
results it is necesary to determine to which type our galaxies, based on their
S\'ersic indices, correspond. 

As we indicated above, \cite{andr95} show that there is a correlation between
the S\'ersic index and morphological type. However, here we need to determine
how these parameters correlate for our dataset. For doing so, we follow
\cite{fasa00} and create a set of simulated galaxies. Their magnitudes,
scale-lengths, position angles and ellipticities are extracted from the
previously described morphological catalogue of 241 galaxies. For each entry
in the morphological catalogue 20 galaxies were simulated, each one with a
different bulge-to-disk ratio, ranging randomly from 0.01 (almost a pure disk)
to 1000 (almost a pure spheroidal).  Each galaxy image was created using {\tt
  galfit}, with a two components S\'ersic model, one with $n_s$=4 for the
bulge and another with $n_s$=1 for the disk, convolving them with the same PSF
used in the morphological analysis. Finally, {\tt noise} was added following
the prescription presented in \cite{coe06} and \cite{sanc04c} for simulating
ACS images, to resemble as much as possible the original F850LP-band image.

The same morphological analysis described in the previous section was
performed for each of the simulated galaxy images, fitting them to a single
component S\'ersic model.  Finally we end up with $\sim$5000 simulations
comprising the recovered S\'ersic index for a certain bulge-to-disk ratio.
\cite{simi86} show the typical bulge-to-disk ratios for the different families
of galaxies in the revised Hubble sequence. In particular, they present the
mean fractional luminosity of the spheroidal component of the galaxies and its
corresponding magnitude difference for each galaxy type.  Figure \ref{fig:sim}
shows the S\'ersic indices of the simulated galaxies as a function of these
magnitude differences ($\Delta$mag). The differences for various families of
galaxies along the Hubble sequence are indicated \cite[extracted from
][]{simi86}. Based on this figure we adopted the following scheme to
morphologically classify galaxies based on the S\'ersic indices: galaxies with
$n_s>$2.5 are considered spheroidals (E-type), galaxies with 1.75$<n_s<$2.5
are considered intermediate type (mostly S0's, but strongly contaminated by
other families of bulge-dominated spirals, like Sa and Sb), and galaxies
with $n_s<$1.75 are considered spirals (L-type). 

This classification scheme is based on the fair correlation between the
bulge-to-total luminosity ratio and the hubble type of a galaxy \cite{simi86}.
However we may admit that some degree of contamination is expected among the
different morphological subgroups due to the dispersion in this correlation.
Although we can clearly distiguish between pure spheroidals (E-type) and
spirals (L-type), the subgroup of intermediate type is most probably less
accurately defined, with considerable contaminations from both extreme
subgroups. Another source of criticism about the adopted scheme is the fact
that some dwarph elliptical galaxies shown low S\'ersic indices
\cite[e.g.,][]{bara03}. Although in galaxy surveys without morphological
pre-selections most of the galaxies with low S\'ersic indices are spiral
galaxies \cite{andr95}, the contamination from dwarph ellipticals could be a
potential problem in the study of galaxy clusters. To address this problem we
perform a visual inspection of the 29 galaxies with $n_s<$1.75 listed in Table
\ref{tab:phot} and \ref{tab:morph} (ie., the ones classified as spirals) using
the previous described HST/ACS images. Only 2 of these 29 galaxies ($\sim$7\%)
could be classified as dwarph spheroidals based on this visual inspection.
However both of them are nearby brighter galaxies, and their low surface
brightness areas (like disk structures) are not clearly distinguished.  In any
case, the possible contamination by dwarph spheroidal galaxies is expected to
be marginal. The remaining galaxies show clear evidence of disk structures.
Taking into account these caveats we consider that the addopted classification
scheme is valid within the context of this study.

\subsection{Spectra extraction}
\label{sec:3}

To deblend and extract the integrated spectra of each individual galaxy in the
field we used a technique for 3D crowded field spectroscopy developed by
ourselves \citep{sanc04a,garc05,sanc06a}. The technique is an extension to IFS
of {\tt galfit} \citep{peng02}, that we named {\tt galfit3d}. It consists of a
deblending of the spectrum of each object in the datacube by fitting of
analytical models. IFS data may be understood as a set of adjacents
narrow-band images, with the width of the spectral pixel. For each narrow-band
image it is possible to apply modelling techniques developed for 2D imaging,
like {\tt galfit}, and extract the morphological and flux information for each
object in the field at each wavelength. The spectra of all the objects is
extracted after repeating the technique through the datacube.  We have already
shown that the use of additional information to constrain the morphological
parameters increases the quality of the recovered spectra \citep{sanc06a}.
With that purpose we used the morphological parameters previously derived from
the HST/ACS F850LP-band image.

\begin{figure}
\centering
\includegraphics[angle=270,width=8cm]{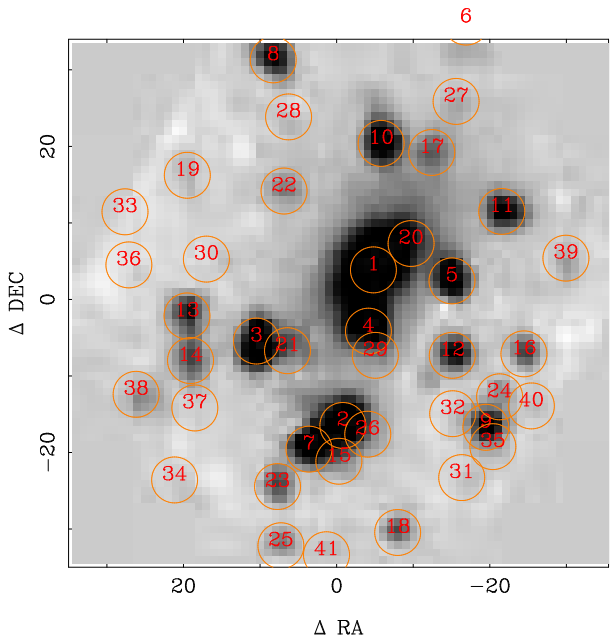}
%
%
\caption{Grayscale representation of the 2D image obtained by coadding
  the flux from the datacube through all the coveraged wavelength
  range (4650-8000 \AA). The circles indicate the location of the
  detected galaxies in the datacube, together with a number indicating
  the label code used to identify them. }
\label{fig:2}       
\end{figure}

We create a 2D image by coadding the flux through the usefull wavelength range
(4650--8000 \AA) covered by our datacube. After this, we cross-checked by eye
this image, looking for those galaxies detected in the F850LP-band for which
we had performed a morphological analysis. We detect 41 objects (40 galaxies
and 1 star) in the collapsed 2D image, out to $z<$ 23 mag. The gravitational
arc-lenses were excluded since their study is out the scope of the current
work. Figure \ref{fig:2} shows a greyscale of the coadded 2D image derived
from the datacube, together with an identification label at the location of
each of the detected objects.

As we mentioned before, we use the previously derived morphological parameters
for each galaxy to extract its integrated spectrum by modelling each galaxy in
the datacube with {\tt galfit3d}. We use the same model that was used to fit
the F850LP-band image, with all the morphological parameters fixed, and
fitting only the flux at each wavelength. For each wavelength, we created a
PSF using postage stamp images of the only field-star in the field-of-view of
our IFS data. As in the previous morphological analysis, the 3D fit was
performed in a sequential way. We first extracted the spectrum of the
brightest galaxy, the central cD, masking the neighbouring galaxies. Once
extracted, its 3D model was subtracted to the original datacube, and the
spectrum of the second brightest galaxy was extracted. The process continued
iteratively until the faintest galaxy spectrum was extracted. The individual
spectra are shown in Appendix \ref{app:1}.  All of them are background
limited. Thus, the signal-to-noise ratio can be estimated from the
background-noise level. We used the variations in the sky spectrum in each
sky-fiber to determine the level of background noise. In this way, we
estimated that the signal-to-noise of the extracted spectra range between
$\sim$1000 and $\sim$5 per spectral pixel for the brightest and faintest
targets. The signal-to-noise ratio is affected strongly by the sky-emission
lines (see Fig.  \ref{fig:sky}), where it drops to less than $\sim$3$\sigma$
even for the brightest targets, respectively. The effect is more severe in the
wavelength range $\lambda>$7200\AA\, where many sky-emission lines happen to
be blended.

The procedure relies on the assumption that the morphological parameters do
not change substantially as a function of the wavelength, something which it
is not completely true, since color gradients are expected in all the
galaxies. But, if they change, we expect this effect to have a minim impact in
the photometry derived with {\tt galfit}.  Thus, if we use the morphological
parameters derived by applying {\tt galfit} to the F850LP-band and we force
the fitting procedure to use them for deriving the photometry at any other
band, the recovered flux should be basically the same as when we let {\tt
  galfit} fit these parameters freely.  Although our previous experiments
shown that this is the case \citep[e.g.][]{garc05}, we have also tested this
hypothesis here, as described in Appendix \ref{app:2}. In any case, the method
is not strongly sensitive to this assumption, as described in \cite{sanc06a}.
In principle, it is possible to perform the extraction procedure using the
parameters derived from the morphological analysis as an initial guess, and
letting {\tt galfit} to derive the correct ones for each wavelength. In this
case, the recovered spectra are noiser and in the residual datacube there is
clear evidence of not correctly extracted structure.

\begin{table*}
 \centering
 \begin{minipage}{160mm}
  \caption{Summary of the parameters derived from the spectra}
  \label{tab:red}
  \begin{tabular}{rrlrrrrr}
  \hline
  IFU& SEx.& $z$ & H$\beta$ & Mg$b$ & Age & Met & Mass \\
  ID & ID  &     &          &       & Gyr & $Z$ & 10$^{10}$M$_\odot$\\
  \hline
1  &1   &0.174 &{\it 1.6$\pm$0.1 }&{\it 3.0$\pm$0.1 }     & 9.8$\pm$2.8 &0.019$\pm$0.011 &102.1$\pm$2.6\\
2  &512 &0.179 & 2.1$\pm$0.1 &3.4$\pm$0.1                 &12.0$\pm$3.4 &0.025$\pm$0.014 & 74.6$\pm$2.8\\
3  &501 &0.163 & 1.1$\pm$0.1 &4.2$\pm$0.2                 &12.9$\pm$3.6 &0.046$\pm$0.026 & 39.5$\pm$2.0\\
4  &554 &0.175 & 2.8$\pm$0.1 &3.6$\pm$0.1                 &12.1$\pm$3.4 &0.025$\pm$0.014 & 67.9$\pm$3.3\\
5  &555 &0.181 & 1.4$\pm$0.1 &4.1$\pm$0.2                 &11.6$\pm$3.2 &0.032$\pm$0.018 & 36.2$\pm$2.2\\
6  &2   &0.181 & 1.5$\pm$0.6 &6.1$\pm$0.9                 &12.7$\pm$3.6 &0.045$\pm$0.025 & 75.7$\pm$4.6\\
7  &509 &0.184 &{\it 1.0$\pm$0.1 }&{\it 3.0$\pm$0.1 }     & 9.9$\pm$2.8 &0.019$\pm$0.011 & 17.2$\pm$1.2\\
8  &570 &0.184 &{\it 1.5$\pm$0.2 }&{\it 2.3$\pm$0.7 }     &12.7$\pm$3.6 &0.045$\pm$0.025 & 39.9$\pm$2.7\\
9  &599 &0.163 & 1.5$\pm$0.1 &4.4$\pm$0.2                 & 9.6$\pm$2.7 &0.019$\pm$0.011 & 10.5$\pm$0.8\\
11 &658 &0.172 & 2.2$\pm$0.1 &3.9$\pm$0.1                 & 9.5$\pm$2.7 &0.018$\pm$0.010 & 11.5$\pm$0.8\\
12 &596 &0.174 & 1.4$\pm$0.1 &4.9$\pm$0.1                 &12.4$\pm$3.5 &0.024$\pm$0.013 & 13.6$\pm$1.2\\
13 &462 &0.179 & 3.2$\pm$0.1 &2.7$\pm$0.2                 & 4.5$\pm$1.3 &0.018$\pm$0.010 &  7.8$\pm$0.6\\
14 &449 &0.178 & 2.7$\pm$0.1 &2.8$\pm$0.2                 & 5.2$\pm$1.4 &0.009$\pm$0.005 &  4.5$\pm$2.3\\
15 &511 &0.162?&{\it 1.0$\pm$0.2 }&{\it 2.5$\pm$0.2 }     & 9.5$\pm$2.7 &0.001$\pm$0.001 &  2.2$\pm$0.3\\
16 &643 &0.167 &{\it 1.2$\pm$0.1 }&{\it 0.7$\pm$0.3 }     & 9.7$\pm$2.7 &0.019$\pm$0.011 &  4.1$\pm$0.5\\
17 &632 &0.179 & 2.8$\pm$0.3 &3.0$\pm$0.2                 & 3.5$\pm$1.0 &0.009$\pm$0.005 &  2.1$\pm$0.3\\
18 &549 &0.172 &{\it ---         }&{\it 3.5$\pm$0.3 }     & 5.2$\pm$1.4 &0.009$\pm$0.005 &  2.8$\pm$0.4\\
20 &560 &0.180 &{\it 2.2$\pm$0.2 }&{\it 0.7$\pm$0.4 }     & 9.7$\pm$2.7 &0.019$\pm$0.011 &  1.8$\pm$0.3\\
21 &533 &0.171 &{\it 1.2$\pm$0.3 }&{\it 1.8$\pm$0.3 }     &11.5$\pm$3.2 &0.031$\pm$0.018 &  4.5$\pm$0.7\\
22 &603 &0.171 &{\it ---         }&{\it 2.7$\pm$0.3 }     & 5.8$\pm$1.6 &0.003$\pm$0.002 &  1.0$\pm$0.2\\
24 &619 &0.159 & 3.1$\pm$0.2 &2.8$\pm$0.5                 &11.4$\pm$3.2 &0.006$\pm$0.003 &  1.5$\pm$0.2\\
25 &484 &0.174 &{\it 1.2$\pm$0.5 }&{\it 0.9$\pm$0.3 }     & 4.2$\pm$1.7 &0.007$\pm$0.004 &  1.0$\pm$0.2\\
26 &513 &0.180 & 3.8$\pm$0.4 &---                         & 7.1$\pm$2.0 &0.019$\pm$0.011 &  2.1$\pm$0.5\\
27 &659 &0.168 & ---         &---                         &12.8$\pm$3.6 &0.045$\pm$0.025 &  1.8$\pm$0.6\\
28 &575 &0.168 & ---         &---                         &12.7$\pm$3.6 &0.044$\pm$0.025 &  3.6$\pm$0.9\\
29 &574 &0.168 & 2.3$\pm$0.7 &---                         & 9.7$\pm$2.7 &0.008$\pm$0.004 &  0.6$\pm$0.2\\
38 &431 &0.170 &{\it ---         }&{\it 1.2$\pm$0.4 }     & 0.3$\pm$0.1 &0.007$\pm$0.004 &  0.1$\pm$0.1\\
41 &507 &0.168 & ---         &---                         & 5.8$\pm$1.6 &0.007$\pm$0.004 &  0.6$\pm$0.1\\
  \hline
\end{tabular}

Uncertain values are written in italic.
\end{minipage}
\end{table*}

\subsection{Redshift determination}

We determined the redshift of each detected galaxy in the field-of-view of our
IFS data by comparing its spectrum with syntethic models for single stellar
populations obtained from \cite{bc04}. We will describe later in detail 
which models were used. We manually tuned the redshift of the model matching
the different absorption features, including the Balmer lines (H$\beta$,
H$\gamma$, H$\delta$), the different Fe and Mg lines covered by our
spectral range, as well as spectral features, like the 4000\AA\ break. Due to
the spectral resolution of our IFS data, and the signal-to-noise of our
targets, we do not expect to have an accuracy better than 0.001 in the
redshift determination. We derived realiable redshifts for 31 galaxies, 28 of
them at the redshift of the cluster ($z = 0.17 \pm$0.2), one is a foreground
galaxy ($z = 0.104$), and other two are backgroud galaxies (both at $z \sim
0.426$).
Table \ref{tab:red} lists the derived redshifts together with the
identifications shown in Fig. \ref{fig:2} (that we will use throughout this
paper) and the {\tt SExtractor} identification listed in Table \ref{tab:morph}
to allow for direct comparison.

Eleven of the 28 galaxies were studied before by \cite{borg92}. The redshifts
derived from our data are coincident with the reported in this article within
the expected errors ($\sigma_z\sim$0.001). We derive new redshifts for 17
galaxies, all of them compatible with the redshift of the cluster. These new
galaxies have been included in the Table \ref{tab:morph}.  In total, the final
sample of cluster members (28) is comparable with that one of previous similar
studies, but with the advantage that no sample has been selected: our sample
is complete within the field-of-view of our IFS data down to our detection
limit.  Of these galaxies, only 6 have been previous studied with enough
quality spectroscopy to derive the properties of their stellar populations
\citep{zieg01}. Therefore, our sample will increase significantly our
knowledge of the core of the cluster.

\subsection{The color-magnitude diagram}

Figure \ref{fig:CM} shows the rest-frame $B$-$V$ vs. M$_V$ color-magnitude
diagram for those cluster members listed in Table \ref{tab:phot}, derived from
the HST/ACS based photometric data described above. The absolute magnitudes
were corrected for cosmological effects and we applied a $k$-correction based
on the same SSP synthetic model. For each synthetic SSP model (described
below) we derived the expected $B$-$V$ colors at the average redshift of the
cluster and in the rest-frame, deriving at the same time the $V$-band
$k$-correction.  An average relation between the observed and rest-frame
colors and the $k$-correction was derived, and applied to the data shown in
Fig.  \ref{fig:CM}.  Red solid circles indicate galaxies with S\'ersic index
$n>$2.5 (E-type), while green stars indicate those with $n<$1.75 (L-type).
Orange solid squares indicate galaxies with S\'ersic indices in between those
values, corresponding to intermediate galaxies. The cD galaxy is
indicated with an orange open cross, being the brightest object in the field.
We prefer to label it with a different symbol since it is well known that cD
galaxies do not follow a S\'ersic profile of any index, showing a flatter and
more extended light profile in the outer regions \citep[e.g.][]{oeml76}. Open
circles indicate the galaxies within the field-of-view of our IFS, with
redshifts listed in Table \ref{tab:red}. The big open square indicates the
only galaxy with a high Asymmetry index.

\subsubsection{Morphological segregation}

To our knowledge this is the first time that a morphologically segregated
color-magnitude diagram is shown for confirmed cluster members.  In previous
works (e.g. Smail et al. 2001) only cluster galaxy candidates have been shown.
The morphologies, based on the S\'ersic indices listed in Table
\ref{tab:morph}, break down as: 11 E-type, 19 intermediate and 29 L-type. If
we consider as E-type galaxies to be all the objects with $n>$1.75, they
represent only a $\sim$50\% of the total. For the complete subsample of
galaxies covered within the IFU field-of-view, the morphologies breaks down
as: 7 E-type, 8 intermediate and 13 L-type. Contrary to previous results
\citep[e.g.][]{zieg01,smai01} this indicates that the cluster population is
not dominated by early-type galaxies, and there is an almost parity between
early- and late-type galaxies.

Our approach to derive the morphologies is substantially different to in
previous attempts, based on visual classifications. There is a considerable
debate over the reliability of visual classifications of early-type galaxies
\citep[e.g.][]{dres97,andr98,fabr00}, and the ability to compare them with
classifications based on profile analysis, like concentration and/or S\'ersic
indices. However, since there are no previous similar studies, we must compare
with the existing ones. \cite{zieg01} show visual morphological
classifications of 19 of their 48 cluster members, based on WFPC/HST imaging.
Their classification was performed by W.Couch, who classified them as 8 E, 1
E/S0, 5 S0, 3 SB/0, 1 Sa and 1 Sab. Taking together all the E-type galaxies
(E, E/S0 and S0), they account for a $\sim$73\% of the total sample.
\cite{smai01} presented morphological classification for 81 cluster candidates
(33 with confirmed redshifts), based on the same analysis as \cite{zieg01}.
Their morphologies break down as: 31 E, 3 E/S0, 28 S0, 5 Sa and 14 late-type
spirals. Taking together all the E-type galaxies, they account for a
$\sim$76\% of the total sample. Even taking into account the different methods
used to derive the morphologies, and the difficulties to compare them, it is
clear that both \cite{zieg01} and \cite{smai01} found fewer L-type galaxies
than what we are finding.

\begin{figure}
\centering
\includegraphics[angle=270,width=8cm]{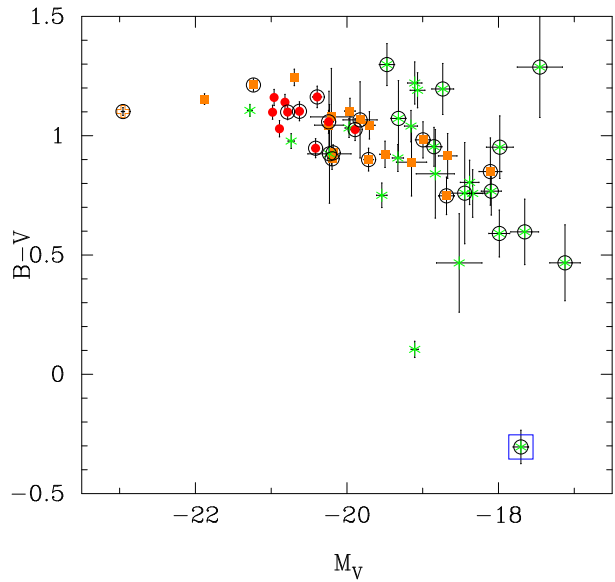}
%
\caption{Rest-frame $B$-$V$ color distribution as a function of absolute
  magnitude M$_V$ for those cluster members listed in Table \ref{tab:phot}.
  Red solid circles indicate galaxies with S\'ersic index $n>$2.5 (E-type),
  while green stars correspond to objects with with $n<$1.75 (L-type). Orange
  solid squares indicate lenticular galaxies, with S\'ersic indices in between
  both values. The cD galaxy is indicated with an orange open cross, being the
  brightest object in the field. Open circles indicate the 28 galaxies within
  the field-of-view of our IFS. The big open square indicates the only galaxy
  with a high Asymmetry index.}
\label{fig:CM}       
\end{figure}

\cite{zieg01} selected galaxies using the previously existing redshift
catalogue from \cite{borg92} and defining a region in the $UBVI$ colour space
which was occupied by cluster members. Finally, only galaxies brighter than
$I<$18.8 mag were targeted by their spectroscopic survey. Although it was
claimed that this selection places negligible restrictions on the stellar
populations of the selected galaxies, and was made with the aim of rejecting
the majority of background galaxies, it may have had an effect on the
morphological segregation. The brightness cut in the $I$-band implies that all
the observed galaxies are brighter than $M_V<\sim$-19.5 mag, asuming an
average $V-I$ color of $\sim$1.5 mag \citep{fuku95}. If we apply such a cut in
our sample (see Fig. \ref{fig:CM}), the morphologies break down as: 9 (7)
E-type, 14 (5) S0's and 11 (2) L-type for the full (IFU) sample. This means a
fraction of $\sim$68\% (86\%) of E-type galaxies, in agreement with the
results from \cite{zieg01}. Therefore, although visual classifications are not
fully reproductible and may produce discrepancies in the morphological
classification of individual objects, they agree with those ones based on
S\'ersic indices on a statistical base.

\cite{smai01} selected 81 cluster candidates based on a deep $K_s$-band survey
($K_s<$19 mag), discarding those objects with clear stellar morphology, in too
crowded areas and/or showing lensed morphologies. Assuming a $z-K_s$ color of
$\sim$3 mag for the reddest objects, based on their own results for $I-K_s$
and an average $z-I$ color of $\sim$0.5 mag \citep{fuku95}, they selected
galaxies brighter than $z<\sim$22 mag, a limiting magnitude similar to that of
our IFU data. Therefore, a difference in the depth of both surveys is not the
reason for the differences in the morphological segregation. We matched their
cluster candidate sample with our full sample (listed in Table
\ref{tab:morph}), finding 39 coincidences. For this subsample of confirmed
cluster members the morphological segregation based on their analysis remains
similar, with a $\sim$82\% of E-type galaxies. Therefore, possible
contaminations from non cluster members are not the cause of the discrepancy.
We examined the S\'ersic indices of those 32 E-type galaxies. Of them, 10 have
S\'ersic indices lower than $n<$1.75, being L-type galaxies following our
scheme, and only 7 have $n>$2.5. Thus, the discrepancy is most probably due to
the different criteria adopted for classifing the objects, illustrating the
risks of direct comparisons between both methods. Statistically speaking
visual classification gives the same distribution of galaxy types as profile
fitting: on average, their E-type galaxies have larger S\'ersic indices
($n\sim$2.6) than their L-type ones ($n\sim$1.7). However, it is well known
that one-to-one cross-checking both methods give different results
\cite[e.g.][]{wolf05}, and profile analysis produces a more realiable and
objetive scheme for classifying objects.

\subsubsection{Analysis of the CM diagram}

The Color-Magnitude (CM) diagram contains extremly useful information to
understand the nature of the cluster population. The $B-V$ color is related to
both age (strongly) and metallicity (weakly) of the dominant stellar
population of the galaxy. On the other hand, the $V$-band absolute magnitude
can be related to the stellar mass. Figure \ref{fig:CM} shows a bimodal
distribution of early- and late-type galaxies in the CM diagram. E-type
galaxies are (mostly) located in the brighter end of the distribution (no one
is fainter than M$_V<\sim-$20.5 mag), covering a narrow range of colors
(1.9$<B-V<$2.2). This is a well known feature of early type galaxies in
clusters and it has been used for decades to select cluster candidates
\cite[e.g.][]{zieg01,sanc02}, and to trace the evolution of the older
population of cluster galaxies \cite[e.g.][]{arag93}.

On the other hand, L-type galaxies are distributed across a wide range of
magnitudes with the brightests almost as red as the early type galaxies, and
the faintests bluer and covering a wider range of colors. For galaxies
brighter than M$_V<-19.7$ mag there are no significant differences between the
core members (galaxies in the IFU field-of-view), and the rest of the galaxies
of the sample. For galaxies fainter than M$_V>-19.7$, it is found that the
cluster core galaxies cover a wider range of colors, with more red galaxies.
However, this difference may be due to a selection effect of the galaxy
samples obtained from the literature. As indicated in Section 4.3 two of the
29 galaxies classified as L-type could be spheroidals on the basis of a
visual inspection of the HST/ACS images. Both of them have red colors,
$B-V\sim$0.9 mag. However they are not the redder L-type galaxies in their
magnitude range.  It is interesting to note here the location of the galaxy
number 38 (IFU$_{\rm id}$), the only one showing a high asymmetry, being the
bluest galaxy of our sample. This galaxy, at the East in Fig. 2 and 3, is
clearly identified as a merger in its lastest stages, showing a clear tail to
the South-West. Its spectrum, Fig.  \ref{fig:spec}, shows strong H$\beta$,
[OIII]4959,5007 and H$\alpha$ emission lines and weaker [NII]6548,6584,
indicating a large amount of star formation. Its blue colors and spectrum also
indicate the presence of an extremely young stellar population, far away from
the global trend of the rest of the cluster members, as we will discuss below.
Contrary to what is expected this galaxy is just at the nominal redshift of
the cluster ($z=0.17$, based on the emission lines), and considering its
projected distance to the cluster center (assumed to be the cD), it is not
likely to be a fly-by object.  To our knowledge it has escaped the attemption
of previous studies, most probably because its morphology and color
automatically exclude it as a cluster candidate. Our IFU data lack the
required spatial and spectral resolution to perform a more detail study of its
properties.

A brief comparison of the color-magnitude distribution of the different kind
of galaxies in the cluster, segregated by their morphology, indicates that
E-type galaxies are more massive and have older and more metal rich stellar
populations. L-type galaxies cover a wider range of masses, ages and
metallicities, with the more massive ones being similar to the E-type
galaxies, and the less massive ones younger and metal poorer. This result
indicates that a bulk star formation process at high redshift may not be the
unique mechanism to explain the population of the galaxies in the cluster. We
will investigate it further in the next sections.

\begin{figure}
\centering
\includegraphics[angle=270,width=8cm]{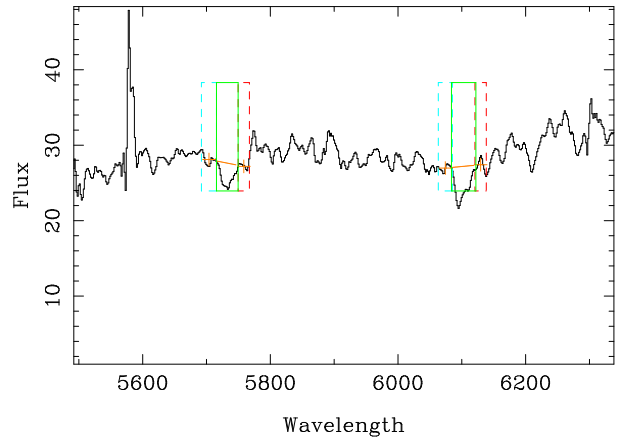}
%
%
\caption{Example of the method used to derive the H$\beta$ and Mg$b$ line
  indices. The plot shows a detail of the spectrum of the galaxy number 2
  (IFU$_{\rm id}$), zoomed in onto the region of interest. For each absorption
  line the boxes indicate the regions were the adjacent continuum are
  estimated (blue and red boxes) together with the region where the absorption
  is measured. The orange line indicates the adopted slope for the continuum.
  The strong feature at 5577\AA\ is the residual of the non perfect
  subtraction of a night sky emission line. }
\label{fig:get}       
\end{figure}

\subsection{Galaxies ages and metallicities}

We obtained spectra of the individual galaxies in the cluster with the aim of
comparing their global spectroscopic properties with their morphological ones
in a relative way, ie., only within the cluster. It is well known that the
spectral energy distribution (SED) of simple stellar populations (chemically
homogeneous and coeval stellar systems) depends on a set of first principles
(e.g., initial mass function, star formation rate, stellar isochrones, element
abundance ratios), to make it possible to generate synthetic stellar
populations from them.  This technique, known as evolutionary synthesis
modeling (e.g.  Tinsley 1980), has been widely used to unveil the stellar
population content of galaxies by reconciling the observed spectral energy
distributions with those predicted by the theoretical framework. Unfortunately
the variation of different physical quantities governing the evolution of
stellar populations produce similar effects in the integrated light of those
systems, leading to a situation in which the observational data is affected by
undesirable degeneracies, like the widely mentioned one between age and
metallicity (e.g.  O'Connell 1976; Aaronson et al. 1978, Worthey 1994a).  The
most widely used technique to compare galaxies with models is to measure
certain line strength indices, on the Lick/IDS index system (Burstein et al.
1984; Faber et al. 1985; Burstein, Faber \& Gonz\'{a}lez 1986; Gorgas et al.
1993; Worthey et al.  1994). People generally try to use a combination of
indices that are most orthogonal in parameter space (e.g. age and metallicity;
although see Cardiel et al. 2003 for a discussion of the impact of random
errors in the selection of the best combination of indices). Following the
Lick/IDS definitions, for our study we have chosen H$\beta$ as a primary age
indicator, and the Mgb index as a metallicity indicator. We use the Mg$b$
index rather than the more conventional [MgFe] one, because both we and
\cite{zieg01} are unable to derive realiable Fe indices for most of the
galaxies. This diagram has the drawback that the models based on solar
abundance ratios do not match the Mg/Fe ratio observed in early-type galaxies
\citep{thom99,thom04}. It is also well known that the ages derived from the
grid are clearly uncertain in an absolute sense. However, we are primarily
interested in a relative comparison of the ages and metallicities of the
galaxies in the cluster, and in this sense the diagram is valid.

As we will discuss below, a proper estimate of those indices requires high
signal-to-noise spectra, and may be strongly affected by the instrumental and
systemic velocity dispersions, the presence of weak emission lines that fill
the absorptions (specially important in the case of H$\beta$), and the
presence of high residuals corresponding to the removal of nearby sky emission
lines.

\begin{figure}
\centering
\includegraphics[angle=270,width=8cm]{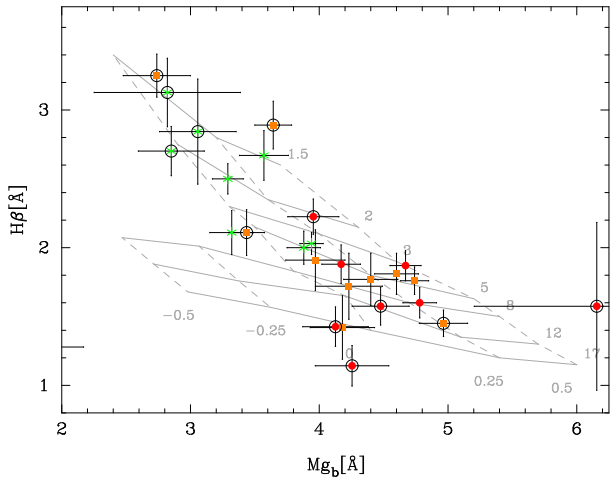}
%
%
\caption{Age - metallicity diagnostic diagram using H$\beta$ as a primary age indicator and Mg$b$ as a metallicity indicator. The symbols are similar to those ones in Fig.\ref{fig:CM}. 
Overplotted is a grid of SSP models from Worthey (1994b). 'Horizontal' lines
follow constant age (Gyr), 'vertical' lines constant metallicity ([M/H]).
Each line is labeled with its corresponding age/metallicity.}
\label{fig:Hb_Mg}       
\end{figure}

\subsubsection{Analysis of line-strength indices}

We derive the H$\beta$ and Mg$b$ equivalent widths for our IFU sample using
the bandpass definitions from the Lick index system revised by \cite{trag98},
shifted to the redshift of each object. We used our own routines and checked
the results with {\tt
  indexf}\footnote{http://www.ucm.es/info/Astrof/software/indexf/}. Figure
\ref{fig:get} illustrates the procedure, showing a detail of the spectrum of
galaxy number 2 and the actual bandwidths used to derive each of the indices.
The derived results are listed in Table \ref{tab:red}, together with the error
estimates (computed as explained below).  We only list the positive values,
since a negative equivalent width means emission, and not absorption.  In most
of the cases these negative values are found in faint targets, and are
simply due to low signal-to-noise of the spectra.  However, in a few cases they
indicate real emission lines, like the strong emission lines of galaxy number
38 (the merger) or fainter ones, like in galaxies numbers 18 and 22. In these
three cases it is possible to derive realiable Mg$b$ indices, but unreliable
H$\beta$ ones. On the other hand, there are a few objects (16, 20, 21 and 25),
with too low Mg$b$ equivalent widths for their H$\beta$ ones. A visual
inspection of their spectra, and a trace back to the raw data shows that they
show cross-talk contamination from ThAr emission lines obtained in adjacent
calibration fibers, which unfortunally lies in the regime of the Mg$b$
absorption. For other four galaxies (number 1, 7, 8 and 15) both H$\beta$
and Mg$b$ seem to be too low. Galaxy number 1 is the cD, a kind of galaxy with
well known deviations of the metacillity sensitive indices
\citep[e.g.][]{nico98}. A visual inspection of its spectrum in the H$\beta$
region indicates possible emission that fill the absorption. A similar
inspection of the spectra of galaxies number 7 and 8 shows clear
contaminations of sky emission line residuals in the adjacent regions to the
H$\beta$ absorption band.  Finally, galaxy number 15 has an uncertain redshift
determination, which strongly affects the derived indices. All these objects
have been excluded from further analysis based on line indices. Of the 6
galaxies in common with \cite{zieg01}, 5 have line-strength indices in common.
In both cases, the strengths of H$\beta$ and Mg$b$ agree within $\pm$0.6\AA,
which is far beyond the estimated errors of both studies. However such
differences are expected on the basis of the different methods used to derive
the strength of the line indices that may strongly affect the results.

We estimated the errors using Monte Carlo simulations of each spectrum, in the
following way: first we derive for each spectrum the best matched single
stellar population model, as we will describe in the next section. This model
is included in Fig.  \ref{fig:spec}. Then, we substract the model from the
spectrum, and derive the absolute value of the difference for each wavelength.
Finally, we smooth this {\it residual} spectrum with a 10 pixels width
median-box filtering. This smoothing was performed to reproduce as much as
possible the noise cross-correlation pattern between adjacent pixels. The
width was selected to match the spectral resolution, since we consider that
this is the range of pixels where the cross-correlation of the noise is
stronger.  
We consider that this rather complex method to estimate the errors is a better
representation of the variance than the much simplier approach
that would be to create a variance spectrum based on the estimated accuracy of
our spectrophotometry (Sec. 2.1 and Appendix B). The reason for that is that
using this latter approach we would not take local effects in the variance,
like unperfections in the sky subtraction, cross-talk contaminations and
unaccuracies on the reduction process. Using the adopted {\it variance} and
our original spectrum we created 100 simulated spectra. The standard-deviation
of the resulting equivalent widths are assumed to be the errors of our
measurements.  Line indices, for galaxies with velocity dispersions larger
than 200 Km/s (derived from the fitting to SSP models, as described below),
are finally corrected as in Jablonka, Gorgas and Goudfrooij (2002) by using
similar coefficients (Jablonka, private communication).

%

The final useful dataset comprises line indices for 26 objects, 12 of them
corresponding to our IFU sample.  Figure \ref{fig:Hb_Mg} shows the Age -
metallicity diagnostic diagram using H$\beta$ as a primary age indicator and
Mg$b$ as a metallicity indicator.  The symbols are similar to those in
Fig.\ref{fig:CM}, including both our determined equivalent widths and those
obtained by \cite{zieg01}. Overplotted is a grid of SSP models from
\cite{wort94}. 'Horizontal' lines follow constant age, 'vertical' lines
constant metallicity. Even though there are in the literature more recent
evolutionary synthesis model predictions suitable for our purposes (e.g.
Bruzual \& Charlot 2003, Vazdekis et al. 2003, Thomas, Maraston \& Kron 2004),
here we are using the classical Worthey (1994b) models because we are just
interested in a differential analysis of our data (i.e. the absolute derived
ages and metallicities should not be taken literally) and we can perform a
direct comparison with the results published by Ziegler et al. (2001).
\cite{zieg01} already noticed that E-type galaxies in this cluster cover a
wide range of parameters within Fig \ref{fig:Hb_Mg}, much wider than similar
samples in the local universe \cite[e.g.][]{kunt01}.  We confirm this
conclusion, since the new galaxies have increased the range of covered
parameters.  The bulk of the galaxies show lower metallicities and younger
ages at a fixed metallicity compared to the ones in the local universe. They
suggested that these differences may be due to variations in the average
masses of the two galaxy samples, in the sense that more massive galaxies
tend to be older.

The \cite{zieg01} sample lacks enough L-type galaxies to draw a clear
conclusion on the dependence of these parameters on the morphology.  With the
new sample it is possible to see that L-type galaxies are found towards
younger ages, on average, than E-type galaxies at any fixed metallicity. On
the other hand they seem to cover the same range of metallicities at any fixed
age. These conclusions have to be taken with care, since most of our L-type
galaxies are, at the same time, less luminous and therefore less massive than
our E-type ones (Fig.  \ref{fig:CM}).  Therefore, if there is a correlation
between the age and the mass, as
suggested by \cite{zieg01}, 
this trend may affect both the E-type and the L-type galaxies, and explain our
results.  We will further investigate this possibility below.


\subsubsection{Fitting to Single Stellar Population models}
\label{sec:fit_SSP} 

\cite{nico03} already showed how sensitive the derivation of physical
parameters from line indices is to the signal-to-noise of the spectra. They
concluded that using the full spectrum, when possible, one can 
derive better parameters, for a similar signal-to-noise. In the previous
section we have seen that even for high signal-to-noise spectra local
contaminations from emission lines, both from the object and/or residuals from
the sky or cross-talk from calibration fibers can strongly affect the derived
indices.

To extract the maximum information for the maximum number of objects we fitted
each spectrum from our IFU sample with synthetic SSP models. Models were
created using the GISSEL code \citep{bc04}, assuming a \cite{salp55} IMF (the
use of a \cite{chab03} will not modify the results), for different ages and
metallicities. We create 72 models covering a discrete grid of 12 ages (5 Myr,
25 Myr, 100 Myr, 290 Myr, 640 Myr, 0.9 Gyr, 1.4 Gyr, 2.5 Gyr, 5 Gyr, 11 Gyr,
13 Gyr, and 17 Gyr), and 6 metallicities ($Z=$0.0001, 0.0004, 0.004, 0.008,
0.02 and 0.05). Each spectrum was then fitted to each of the 72 models using
FIT3D \citep{sanc06b}. This code resamples the model to the resolution of the
data, convolve it with a certain velocity dispersion, shift it to the redshift
of the target and scale it to match the dataset by a $\chi^2/\nu$ minimization
scheme. Therefore, for each SSP model, these are the three free parameters in
the fitting process: velocity dispersion, redshift and a scaling factor
related to the luminosity. The redshift of the object was fixed to reduce the
degrees of freedom. Only the spectral region at wavelengths bluer than 7200
\AA\ was used in this analysis. At redder wavelengths the strong night sky
emission-lines and the telluric absorptions have not properly been corrected
for in our reduction process, and strong residuals are seen in the spectra.
Wavelength regions possibly affected by imperfect sky emission line
subtraction were also masked.  We derived the velocity dispersions for each
object spectrum in the fitting process, including both the instrinsic and
instrumental ones. We cannot perform an accurate correction of the
instrumental effects cause we lack of the proper comparison observations of
velocity calibration stars during the night. This prevents us to use them as a
mass indicator \citep[e.g.][]{zieg01}, using them only to correct the
line-indices strength, when required. The corrected valocity dispersions range
between $\sim$100-400 km s$^{-1}$, and despite the lack of accuracy of our
derived velocity dispersions they agree with the ones derived by
\cite{zieg01}, for the 6 objects in common.

\begin{figure}
\centering
\includegraphics[angle=270,width=8cm]{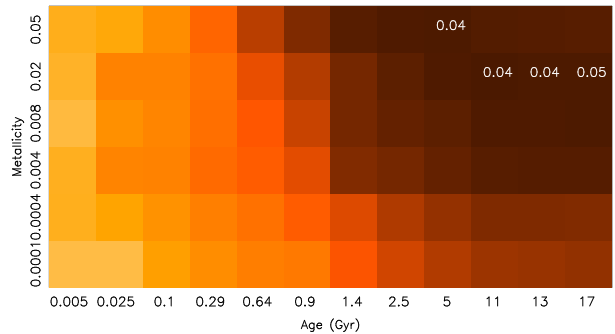}
%
%
\caption{Greyscale image of the 2D distribution of $\chi^2/\nu$ derived from
  the fitting of a single galaxy spectrum to a grid of SSP models of different
  ages and metallicities. Dark colors indicate lower $\chi^2/\nu$ (ie., better
  fits), while light colors indicate higher vales (ie., worse fits). The
  $\chi^2/\nu$ of the four best fits are labeled. The plotted example
  corresponds to the second brightest galaxy in our field-of-view (id=2 in Table \ref{tab:red})}
\label{fig:fit}       
\end{figure}


We obtain for each spectrum a grid of 72 $\chi^2/\nu$, each one corresponding
to an age-metallicity pair model. Figure \ref{fig:fit} illustrates the result
of the fitting technique. It shows a greyscale of the $\chi^2/\nu$ map derived
after fitting the spectrum of galaxy number 2 to the grid of models, each one
corresponding to an age-metallicity pair. Similar results are found for the
rest of the objects. In the fitting procedure we assumed a simple
root-mean-square of the flux as an estimation of the uncertainties in the
spectra, to derive the $\chi^2/\nu$. Therefore these values are not valid to
compare the results of the fit for different objects, and just to perform
internal comparisons between the results of the fit for the same object. The
$\chi^2/\nu$ of the four best fits are indicated in the figure. In general it
is not possible to distinguish which is the best fitting model among the best
four ones.  Therefore, we estimate the age and metallicity of each object by
averaging the ages and metallicities of the four best models, weighted by the
inverse of the corresponding $\chi^2/\nu$ values.  Errors for the estimated
parameters are derived by Monte Carlo simulations: for each spectrum we
estimate the {\it variance} spectrum as described above.  Then we created 30
simulated spectra, using as input the original spectrum and adding noise
following the estimated {\it variance}. For each of the 30 simulated spectra
we run the fitting process over the 72 models (i.e., 2160 fits per input
spectrum), and the age and metallicity are derived following the recipe
described above. The standard deviation of the recovered distribution of ages
and metallicities for the simulated spectra is considered as a good
estimate of the error of the measured parameters for the input spectrum. The
final errors for the ages and metallicities are computed coadding
quadratically these estimated errors to the standard deviation of the values
for the four best models. The results from this procedure are listed in Table
\ref{tab:red}.


The method relies on certain assumptions: (i) we consider that the use of the
full spectrum is a better method to recover the age and the metallicity of the
main stellar population than just using the line indices and (ii) we consider
that our spectrophotometric calibration is stable from blue-to-red, i.e., that
the shape of the spectra is well recovered. This is a sensitive assumption,
since the whole method relies on it. To test its validity we compared our
integrated spectra with the available broad-band photometry (Fig.
\ref{fig:spec}). The photometric dataset, derived from the HST/ACS data,
covers the full optical wavelength range, from $\sim$4700\AA\ to
$\sim$9000\AA, and in particular the wavelength range of our spectroscopic
data. In most of the cases our spectroscopic data match pretty well with the
broad-band photometric ones.  Even more, the extrapolation of the data using
the best fitted SSP synthetic model reproduces also the expected photometry in
wavelength ranges not covered by our spectroscopic data in most of the cases.
Only galaxies number 9, 11, 16, 18, 22 and 24 show a slight missmatch
($\sim$2$\sigma$) between the extrapolated flux and the z$_{850}$-band
photometry.  If we take into account that the spectrophotometry seems to match
for the rest of the bands and for the rest of the objects, we have strong
arguments to consider that is also valid for thes ones: We used an IFU, and
therefore the flux calibration was performed only once for the full dataset.


Finally we assume that (iii) the adopted grid of models comprises a good
representation of the expected stellar populations. We tried not to pre-judge
the composition of the stellar population and cover the wider possible range
of ages and metallicities. On the other hand, we must limit the number of
input models for a reasonable performance of the fitting procedure. This
imposes limitations on the sampling of the grid, and therefore, in the
accuracy of the recovered age and metallicities. We must take that into
account when drawing conclusions from our modelling.

\subsubsection{The Age-Metallicity Distribution}

\begin{figure}
\centering
\includegraphics[angle=270,width=8cm]{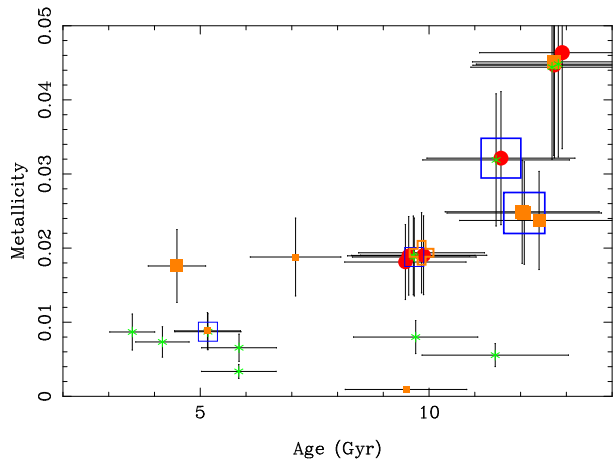}
%
%
\caption{Distribution of the metallicities as a function of age derived by the fitting procedure for the 28 galaxies our IFS sample. The symbols are similar to those ones in Fig.\ref{fig:CM}. The size of the symbols indicate the mass, with larger symbols for more massive objects. All the E-type galaxies ($n>$2.5), and most of the $n>1.75$ ones have stellar populations older than $\sim$9 Gyr and a wide range of metallicities, $0.02<Z<0.05$.}
\label{fig:Age_Met}       
\end{figure}

Figure \ref{fig:Age_Met} shows the distribution of the metallicities as a
function of age derived using the previously described fitting procedure for
the galaxies in our IFU sample. The symbols are similar to the ones in
Fig.\ref{fig:CM}.  The size of the symbols indicates the mass (see below for a
description of how it was estimated), with larger symbols for more massive
objects. The merging galaxy (IFU$_{\rm id}=$38), is not visible in this plot,
since the derived age for its stellar population, $\sim$250Myr, is far from
the average of the rest of the sample. This result is expected due its $B-V$
color, and the strong star formation process that is derived from the strength
of the emission lines on its spectrum. In this particular case a SSP model is
not a good enough to describe the expected mix of populations resulting from a
merging process, with at least an underlying stellar population, and a younger
one resulting from the induced star formation process.

If we compare this figure with the H$\beta$-Mg$b$ distribution shown in Fig.
\ref{fig:Hb_Mg} evident differences can be seen. However, a direct comparison
has to be done with care. First, the absolute ages and metallicities derived
from both figures cannot be compared, since they come from completely
different SSP models. We used the \cite{wort94} models in the previous figure
to be consistent with \cite{zieg01}, while the models used in the fitting
procedure are derived from the GISSEL code \citep{bc04}. However, relative
comparisons between the different galaxies are possible, in principle. One
needs to take into account that in Fig.\ref{fig:Hb_Mg} we only represent 12 of
the 27 objects shown in the current plot, i.e., mostly the brightest ones. On
the other hand, the objects from \cite{zieg01} are not plotted in the current
figure. However, the conclusions about the stellar populations of individual
galaxies derived from both figures agree qualitatively: the galaxies with
young and old stellar populations have young and old stellar populations in
both figures, and the galaxies richer and poorer in metals are richer and
poorer in both figures too.

We consider the current figure a more accurate representation of the real
distribution of ages and metallicities of the galaxies in the cluster (or at
least in the core). The plotted parameters were derived for the IFU sample,
which is complete to our detection limit, without any pre-selection and any
{\it a posteriori} exclusion. In general it is found that the galaxies with
older stellar populations covers a wider range in metallicities, being more
metal rich, on average, than the galaxies with younger stellar populations.
This result does not agree with the trend  between these two
parameter found in field ellipticals, where galaxies with younger stellar
populations are more metal rich \citep[e.g.][]{trag00}. However, our
sample comprises a large fraction of L-type galaxies. As in Fig \ref{fig:CM}
we found a clear different behaviour for the E-type and L-type galaxies. All
the E-type galaxies with $n>$2.5 (spheroidals), and most of the $n>1.75$ ones
(intermediate galaxies) have stellar populations older than $\sim$9 Gyr,
covering a wide range of metallicities, $0.02<Z<0.05$.  On the other hand, the
L-type galaxies ($n<1.75$), cover a much wider range of ages, from $\sim$3 Gyr
to $\sim$13 Gyr (keeping in mind that the absolute ages have no meaning), and
metallicities, $0.001<Z<0.05$. Interestingly, the L-type galaxies with the
largest metallicities ($Z>$0.015) seems to have similar ages as the E-type
ones, and the L-type galaxies with smaller metallicities cover a wider range
of ages.  The most metal rich galaxies are the more massive ones (with larger
symbols), and therefore, this result may indicate that more massive galaxies
tend to be older whatever their morphologies.  We will discuss that in detail
below.  This result is certainly not clear from Fig.\ref{fig:Hb_Mg}.


\begin{figure}
\centering
\includegraphics[angle=270,width=8cm]{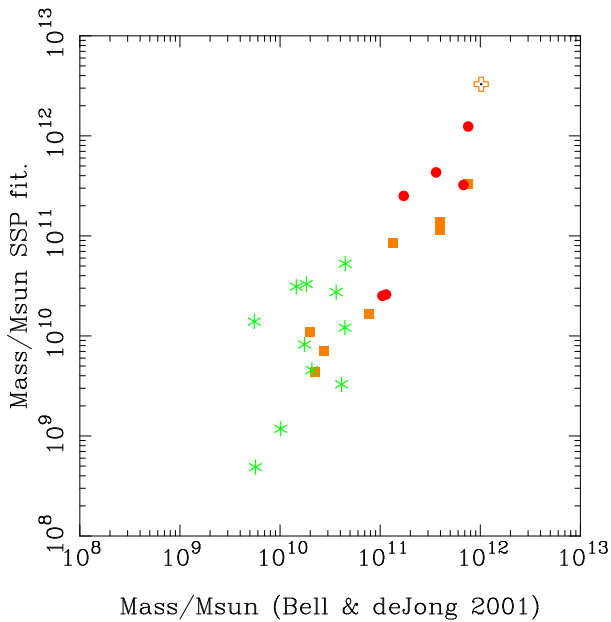}
%
%
\caption{Distribution of the stellar masses for the galaxies of our sample
  derived from the results of the fitting procedure to SSP models along the
  stellar masses derived using the Bell \& de Jong (2001) approximation.}
\label{fig:Mass_Mass}       
\end{figure}


\subsection{Stellar populations and Masses of the Galaxies}

In previous sections we have already seen that the masses of the galaxies are
directly related to the nature of their stellar populations. The
Color-Magnitude diagram, Fig. \ref{fig:CM}, shows that less luminous (and
massive) galaxies show a wider range of colors than more luminous objects. The
comparison of the H$\beta$-Mg$b$ distribution of this cluster with that of
the local universe, including more massive galaxies, makes \cite{zieg01}
think that there might be a trend of metallicities/ages with the mass of the
objects. Figure \ref{fig:Age_Met} shows that more massive galaxies are, on
average, older and more metal rich. We investigate here the possible relations
between the stellar populations and the morphology with the mass of the
galaxies.

\begin{figure}
\centering
\includegraphics[angle=270,width=8cm]{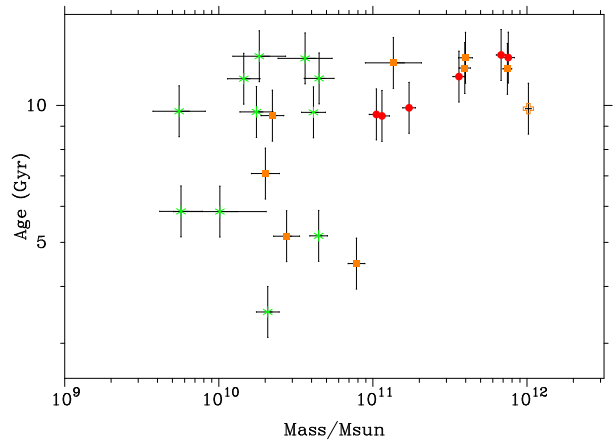}
%
%
\caption{Distribution of ages along stellar masses for the our galaxy sample. The symbols
  are similar to those ones in Fig.\ref{fig:CM}. There is a clear
  morphological segregation. L-type galaxes are less massive than
  10$^{11}$ M$_\odot$, and have a wider range of ages than E-type galaxies.}
\label{fig:Age_Mass}       
\end{figure}

It is well known that the masses of spheroidal or early-type galaxies are
directly related to their stellar velocity dispersions ($\sigma$), due to
their nature. This parameter, directly accesible from the spectra, has indeed
been used as a mass indicator for decades, leading to a set of well establish
scaling relations when comparing it with different properties of the galaxies.
Among these relations the most important ones are the luminosity-velocity
dispersion relation or Faber-Jackson relation \citep{fabe76}, the surface
brightness, effective radius and velocity dispersion relation or Fundamental
plane \citep{dres87,djor87,bend92} and the Mg-velocity dispersion relation or
Mg$b$-$\sigma$ relation \citep{bend93,coll99}.  These scaling relations have
been used to compare between different populations of early-type galaxies at
different redshifts \citep[e.g.][]{zieg01}, to trace the evolution of this
kind of galaxies.  For late-type galaxies similar relations are found when
using the maximum rotating velocity, a parameter directly related with the
mass of the rotating disk.

All these relations have a direct correspondence with relations between the
mass and the different properties of the galaxies: The Faber-Jackson relation
can be understood as a Luminosity-Mass relation, the Fundamental plane as a
Mass-scale relation and the Mg-$\sigma$ as a Metallicity-Mass relation.  Our
spectroscopic data lack the required resolution and signal-to-noise for a
proper determination of the velocity dispersion for most of the galaxies,
therefore, we prefer to investigate these relations using the stellar masses
of the galaxies. This approach has the advantage that we can explore at the
same time all the galaxies of the sample independent of their morphology.

We estimate the stellar masses using the average relation found by
\cite{bell01} between the $M/L$ for the $B$-band luminosity and the $B-V$
color. For doing that we derive the rest-frame $B-V$ colors of our objects
using the best fitted SSP model described before, synthesizing the magnitudes
by convolving the model with the filter responses, and obtaining the $B$-band
absolute magnitude from our photometric data. It is important to note that the
masses of \cite{bell01} are just relative values, calibrated under the
assumption that spiral galaxies are in a maximum-disk situation. However, this
will not affect our results since we are interested in relative comparisons
between different families of galaxies. The derived masses are listed in Table
\ref{tab:red}. The errors listed in the table were derived from the
photometric errors in the $B$-band magnitudes listed in in Table
\ref{tab:phot}, and they do not include any other possible source of error
(like the unaccuracy of the fitting process or the derived synthetic colors).
However, we consider that this error dominates any other possible one.  The
masses of the objects range between $\sim$10$^9$ and $\sim$10$^{12}$
M$_\odot$, being the cD galaxy the most massive galaxy and the merging galaxy
the least massive one.

\begin{figure}
\centering
\includegraphics[angle=270,width=8cm]{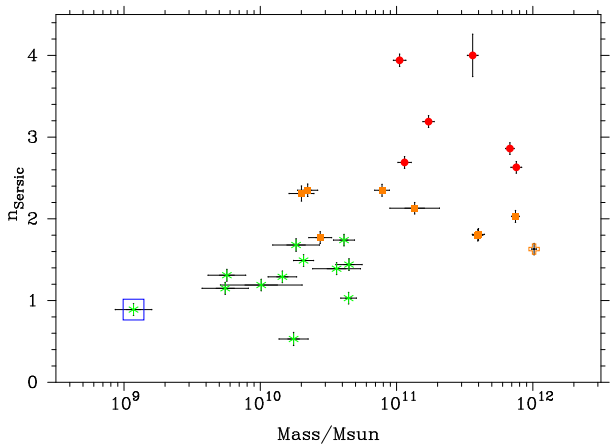}
%
%
\caption{Distribution of S\'ersic indices as a function of the stellar masses for the
  galaxies of our sample. The symbols are similar to those in
  Fig.\ref{fig:CM}. There is a clear trend in the sense that E-type galaxies
  are more massive than L-type ones.}
\label{fig:Nser_Mass}       
\end{figure}

\subsubsection{The Age-Mass relation}

Figure \ref{fig:Age_Mass} shows the age distribution as a function of stellar
masses for the galaxies of our IFU sample. The symbols are similar to those in
Fig.\ref{fig:CM}. The division of galaxies in terms of morphologies is
remarkably similar to that of the Color-Magnitude relation (Fig.\ref{fig:CM}),
although in the current plot the segregation between early- and late-type
galaxies is much stronger. Most of the E-type galaxies ($n>1.75$) and all the
ones with $n>2.5$ are more massive than $\sim$7 10$^{10}$ M$_\odot$ and have
stellar populations older than $\sim$9 Gyr. On the other hand, all the L-type
galaxies are less massive than this limit, and the ages of their stellar
populations spread towards younger ages. There is no obvious trend of age
with mass, but there is a lack of young stellar populations in massive
galaxies.  Since all of them are E-type ones, it is not possible to know if
this deficit is due to the mass or the morphology of the galaxies.  However
such a trend it is expected, at least for field early-type galaxies.  Recently
\cite{pele06} reported a trend between the age and the mass for field
early-type galaxies based on data from the SAURON project \citep{baco01}.

Figure \ref{fig:Nser_Mass} illustrates this effect. It shows the distribution
of S\'ersic indices as a function of the stellar masses for the galaxies of
our IFU sample. The symbols are similar to those in Fig.\ref{fig:CM}.  There
is a strong correlation between both parameters, in the sence that E-type
galaxies are massive and L-type ones less massive. The relation is clearer
when masking the cD and the merger galaxies, for which the profiles are not
well represented by the adopted model.

Similar results were already presented by \cite{zieg01} for the early-type
galaxies of A2218 in the outer regions. They compared the H$\beta$ index with
the stellar velocity dispersion finding that the fainter galaxies shows a
larger scatter in their H$\beta$ values, indicating a wide range of star
formation histories. Our survey goes deeper, sampling galaxies much fainter
(and less massive) that those ones studied by \cite{zieg01}, extending their
conclusions to fainter regimes.


\subsubsection{The Metallicity-Mass relation}

\begin{figure}
\centering
\includegraphics[angle=270,width=8cm]{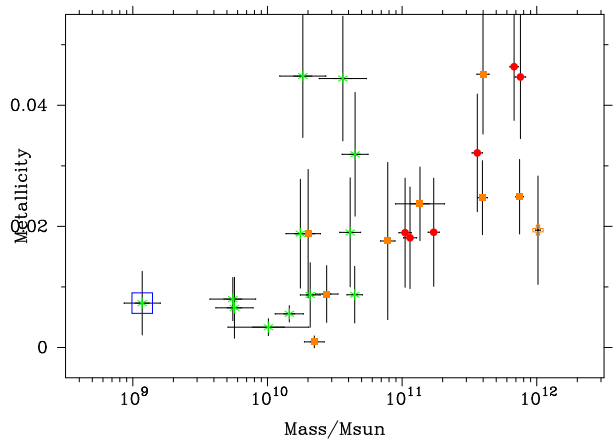}
\caption{Distribution of the metallicities as a function of stellar mass for the
  galaxies of our sample. The symbols are similar to those ones in
  Fig.\ref{fig:CM}. There is a clear correlation between both parameters, in
  the sense that more massive galaxies are more metal rich than less massive ones.}
\label{fig:Met_Mass}       
\end{figure}

As we pointed out above, the Mg$b$-$\sigma$ relation can be understood as a
metallicity-mass relation. Figure \ref{fig:Met_Mass} shows the distribution of
the metallicities as a function of the stellar mass for the galaxies of our
IFU sample.  The symbols are similar to those ones in Fig.\ref{fig:CM}. As
expected there is a clear correlation between both parameters
($r=$0.997,$P>$99.99\%), in the sense that massive galaxies of any kind are
more metal rich than less massive ones. A similar result is found when using
the mass derived by the fitting to SSP synthetic models, instead of the
adopted one, based on the Bell \& de Jong (2001) relation. Although the result
is not surprising for the E-type galaxies, it is interesting that the L-type
galaxies follow a similar relation, but with an offset.
The trend between metallicity and mass seems to be universal
for any kind of galaxy, despite of its morphology.  However, for any fixed
mass the L-type galaxies seem to be more metal rich than the E-type ones. This
effect is not due to a difference in age of the stellar populations, since
at equal age we observe a considerable range in metallicity.  However, our
sample is too reduced to draw firm conclusions.

\section{Discussion}
\label{sec:5}

The galaxy population of clusters in the local universe are dominated by
early-type galaxies (E's and bulge-dominated spirals) \cite{hubb31,oeml74},
with a reduced fraction of late-type galaxies (Spirals, Irregulars and
Mergers). There is a general agreement about the ages of the stellar
populations in early-type cluster galaxies. It is well known that they show a
strong color-magnitude relation at various redshifts consistents with passive
evolution of stellar populations that formed at $z>$2-3, with an slope due to
the mass-metallicity relation \citep[e.g.][]{stan97,stan98,koda98,pogg02}.
The spectrocopic features and the scaling relations, like the
fundamental-plane, the mass-to-light ratio, the Mg$b$-$\sigma$ and the
Faber-Jackson relations, explored at different redshift are also consistent
with passive evolution of an old stellar population
\citep[e.g.][]{bend96,dokk96,kels01,zieg01,dokk01}.

However, this apparently simple scenario for the evolution of early-type
galaxies in clusters needs to be contrasted with another results.
\cite{fasa00} show that there is a substantial morphological evolution in
clusters: as the redshift decreases the S0 population tends to grow at the
expense of spiral population, whereas the frequency of spheroidals remains
almost constant. Therefore the population of early-type galaxies observed in
distant clusters does not necessarily comprise all the early-type galaxies
existing at $z$=0 \citep{pogg02}. On the other hand, the blue galaxies
observed in distant clusters, responsible of the Butcher-Oemler effect, must
largely become red and fainter. There is strong evidence for a quenching of
star formation in galaxies as a consequence of the infall in the cluster dense
regions \citep{pogg01,pogg04}, although a previous starburst may be also
present \citep{pogg04}. For galaxies with a star formation history typical of
L-type, once truncated, it is expected that they become redder and fade by
0.5-1.5 mag in $\sim$2-5 Gyr \citep{pogg04}. Finally, although the
spectrophotometric and the morphological evolution seems to be largely
decoupled, it is expected that after longer timescales there will a
morphological transformation from pure L-type to bulge-dominated
spirals \citep{fasa00}. All these results suggest that the early-type galaxies
in clusters are composed of a mix of different stellar populations, with very
different origins, contrary to the simple scenario drawn above.

How do our results compare with previous knowledge of the evolution of
galaxies in clusters? The morphological distribution of the cluster members
derived from our analysis is: 11 (7) E-type, 19 (8) lenticulars, and 29 (13)
L-type, for our complete sample (our IFU sample). In percentages it is: 19\%
(25\%) E-type, 32\% (29\%) lenticulars and 49\% (46\%) L-type. These fractions
differ strongly with the estimated ones for clusters at similar redshifts:
30\% E-type, 50\% lenticulars and 20\% L-type \citep{fasa00}, showing a much
larger fraction of late-type galaxies. \cite{fasa00} estimated these numbers
based on images with limiting magnitudes of the order of $r\sim$20.2-20.5 mag,
i.e. $\sim$1.5-2 mag less deep than our current survey \citep[taking into
account the range of possible $r-z$ colors for any kind of galaxy at this
redshift, from $\sim$0 mag to $\sim$0.6 mag, ][]{fuku95}.  If we consider only
the objects brighter than $z<$20 mag, the new morphological segregation is: 9
(7) E-type, 16 (7) lenticulars and 16 (4) L-type. These percentages are within
the values reported by \citep{fasa00}: 22\% (39\%) E-type, 39\% (39\%)
lenticulars and 39\% (22\%) L-type. Therefore, there is an increase of the
fraction of late-type galaxies at fainter magnitude regimes.

Looking at Figures \ref{fig:CM} and \ref{fig:Age_Mass} it is possible to
identify two kind of L-type galaxies in the faint magnitude regime: (i) a
group of blue and faint galaxies, with young stellar populations ($<$8 Gyr),
and (ii) another group of red and faint galaxies, with older stellar
populations ($>$8 Gyr). The first ones are the kind of galaxies responsible
for the Butcher-Oemler effect. Galaxies recently captured by the cluster, with
relatively young stellar populations and traces of recent star formation
processes. Due to their colors these galaxies would be detected in surveys in
the optical/blue bands. The second group are the kind of galaxies predicted in
the two-steps evolution hypothesis for galaxies captured by clusters, where
the star formation was quenched first, and then there was a slower
morphological transformation \citep{pogg01,pogg02}. As a first step the
late-type galaxies fall into the cluster, and due to gas stripping the star
formation process is abruptly interupted (maybe after a short starburst
process). Then, these galaxies become fainter and redder, while still showing
a spiral shape.

Clusters with the highest fraction of blue galaxies shows signs of recent
mergers, and there are larger fractions of starforming galaxies in
substructured than in relaxed clusters \citep[e.g.][]{smai98,mete00,pimb02}.
Theoretical simulations predict that the cluster mergers induces a starburst
processes, due to the strong tidal forces involved \citep[e.g][]{moss00}, a
similar effect of galaxy-galaxy interactions \citep[e.g.][]{miho96}.  Abell
2218 shows two clear peaks in the X-ray and galaxies distribution
\citep{mcha90}, indicating that it most probably has suffered a recent (or
ongoing) merging.  This may account for the moderate Butcher-Oemler fraction
of blue galaxies quoted in classical studies \citep[$\sim$10\%][]{butc84}.

Galaxies captured by the cluster may suffer similar tidal forces, which may
induce an enhancement of the star formation process, previous to the quenching
due to gas stripping \citep{pogg02}. The location of the k+a (post-starburst)
galaxies, and the strength of their Balmer absorption features strongly
supports this hyphotesis \citep{pogg04}. To our knowledge, Abell 2218 is the
only cluster at cosmological distances where HI is observed in emission
\citep{zwaa01}, at the location of a spiral galaxy in the outer regions of the
cluster. This galaxy is most probably an infalling spiral, consuming its gas
reservoir with ongoing star formation (estimated SFR of $<$1.4 M$\odot$/yr).
More spectacular is galaxy number 38 from our IFU sample, a blue disrupted
galaxy at the nominal redshift of the cluster with a very little projected
distance from the center of the cluster. It is possible that this galaxy has
been recently captured by the cluster, and that it is moving in an orbit
perpendicular to the line-of-sight. If not, there should be a conspirancy
between its cosmological redshift and line-of-sight velocity to compensate
each other so that it appears to be at the nominal redshift of the cluster,
which is an unlikely explanation. Its emission lines indicate a moderate
amount of ongoing star formation, with a SFR$\sim$6 M$\odot$/yr, derived from
the H$\alpha$ emission \citep{kenn83,kenn94}.  No dust is derived from the
H$\alpha$/H$\beta\sim$ ratio, $\sim$2.9, nearly the nominal value for case B
recombination. Its dominant stellar population has an age of $\sim$250Myr,
which could be a feasible time-scale for the infall process. Its morphology is
similar to that one of a merger in its last stage, however, the strong tidal
forces that may suffer during the infalling process may have produce the same
effect. If the proposed scenario for the evolution of galaxies in cluster is
true, this galaxy would be a good candidate for a progenitor of the observed
k+a galaxies at low-redshift \citep{pogg04}. In a recent study of Abell 2125,
a dense cluster at $z\sim$0.247, \cite{owen06} found a galaxy with strong
emission lines (maybe even with an AGN), blue colors, and a disturbed
morphology. Although that galaxy is brighter than galaxy number 38, both
systems seem to experience the same physical process.

Our results indicate that there is a correlation between the morphology and
the mass, in the sense that less-massive galaxies are mostly L-type and
more-massive ones are mostly E-type (Fig. \ref{fig:Nser_Mass}). This result is
natural in the proposed scenario, where the morphological transformation
between L-type to bulge-dominated spirals, and later to E-type galaxies is a
consequence of consecutive merging processes.  We have also found a clear
correlation between the mass and the metallicity, in the sense that more
massive galaxies are more metal rich (Fig. \ref{fig:Met_Mass}). This
correlation is an extension of the well known Mg-$\sigma$ relation for
spheroidal galaxies, which seems to be independent of the galaxy morphology.
Although we have a very reduced sample, it seems that L-type galaxies are more
metal-rich than E-type ones at any fixed mass. Figure \ref{fig:Met_Mass} also
indicates that the more metal rich L-type galaxies are never more metal than
the E-type ones, being only less massive. To our knowledge this effect has not
been previously reported in the literature, and it may well be an effect of
the very reduce sample. If not, it may imply that the galaxies reach their
maximum metallicity when they are still spirals, maybe due to the enhancement
of the star formation rate when infalling into the cluster, and that they
later evolve to more massive bulge-dominated spirals and E-type galaxies
keeping this metallicity.

\section{Conclusions}
\label{sec:6}

We obtained deep IFS data ($I_{lim}\sim$21.4 mag) covering a field-of-view of
$\sim$74$\arcsec$$\times$64$\arcsec$ centred on the core of the galaxy cluster
Abell 2218, sampling the wavelength range 4650-8000\AA\ with a resolution of
$\sim$10\AA\ (FWHM). Combining these data with a deep HST/ACS F850LP-band
image ($z_{lim}\sim$28 mag), and ground-based and HST/WFPC photometric data, we
have obtained the following results:

\begin{itemize}
  
\item Morphological parameters of all the galaxies in the field-of-view of the
  HST/ACS image down to $z<$22.5 mag were obtained, based on an automatic
  procedure using {\tt galfit}.
  
\item The integrated spectrum of each of the 43 objects detected in the IFU
  datacube was obtained, using a modeling technique for IFS ({\tt galfit3d}),
  which ensures a proper deblending of multiple components in crowded-fields.
  
\item Of the detected objects, 28 were classified as cluster members based on
  their redshifts. They comprise the first volume and flux limited sample of
  confirmed cluster members obtained for the core of this cluster.  We need to
  remark that the sample was obtained without any pre-selection based on
  morphology, colors or magnitudes.
  
\item By combining the different previous redshift and photometric surveys
  with our new morphological information an extended sample of 59 cluster
  members was created, 31 of them sampling the outer regions of the cluster
  (out of the core).
  
\item The color-magnitude diagram of the full sample of confirmed cluster
  members, as a function of the morphology, was analyzed. We have found that
  the fraction of early-type galaxies is lower than previously reported, most
  of them being red and luminous galaxies. On the other hand, the L-type
  galaxies spread over a wider range of colors and luminosities with respect
  to the E-type ones and previously published catalogs.
 
\item There is no significant difference between the Color-Magnitude
  distributions of the galaxies in the core of the cluster and in the outer
  regions if one distinguishes between different morphologies. If any, we
  found a few more L-type faint and red galaxies in the core. However, this
  may be due to the biases of the pre-selection of candidates in the samples
  obtained from the literature (in the outer regions).
  
\item A strongly distorted blue galaxy is found in projection at the core of
  the galaxy cluster. Its redshift corresponds to the nominal redshift of the
  cluster, and therefore, it is more likely that it is physically located in
  the core. A visual inspection of the object indicates that it looks like
  a merger in an advance stage of evolution, dominated by a young stellar
  population. However, if we are observing it in the moment of being captured by
  the cluster, the strong tidal forces that it is suffering may produce the
  observed disruption on its morphology.

\item The analysis of the age and metallicity of a reduced subsample of
  galaxies (24 galaxies, 12 core members) based on line indices indicates that
  L-type galaxies on average spread towards younger ages  than E-type ones. On
  the other hand, they seem to cover the same range of metallicities at any
  fixed age.
  
\item The age and metallicity of each galaxies in the core of the cluster was
  derived by a fitting procedure of its integrated spectrum to single stellar
  population models. The age-metallicity diagram of these galaxies indicates
  too that L-type galaxies cover a clearly wider range of ages, and a
  slightly wider range of metallicities. The more massive L-type galaxies seem
  to have similar ages and metallicities than the E-type ones, while the less
  massive ones tend to be younger and less metal rich.
  
\item The distribution of ages and S\'ersic indices as a function of the
  stellar masses of the core members shows that L-type galaxies are less
  massive on average.  They spread over a wider range of ages, but it is not
  possible to determine if this is due to their morphologies or masses. On the
  other hand, more massive galaxies tend to have older stellar populations
  and they are all early-type galaxies.
  
\item There is a clear trend between the metallicity and the mass, in the
  sense that more massive galaxies are more metal rich. This trend is just a
  change of coordinates of the well known Mg$b$-$\sigma$ relation for the
  early-type galaxies. However, we found that this trend is also present for
  the L-type galaxies, but with an offset: they are more metal rich than
  E-type galaxies at any fixed mass.

\end{itemize}

Our results agree with the proposed two-steps scenario for the evolution of
galaxies in clusters \citep[e.g.][]{pogg02}, in which we have (i) a primordial
population of early-type galaxies formed at early epochs with stellar
populations that are mostly old, and (ii) a new population of L-type galaxies
that are captured by the cluster, infalling, suffering a short enhancement of
the star formation that is later quenched by the interaction with the
environment.  The galaxies then evolve passively, becoming redder and fainter,
while keeping the spiral morphology. The disks dim due to the lack of new
formed stars, and the galaxies become bulge-dominated spirals, after longer
time-scales. Later galaxy-galaxy interactions and dry merging processes build
up new massive spheroidal galaxies.

\section*{Acknowledgments}

Based on observations collected at the Centro Astron\'omico Hispano Alem\'an
(CAHA) at Calar Alto, operated jointly by the Max-Planck Institut f\"ur
Astronomie and the Instituto de Astrof\'sica de Andaluc\'\i a (CSIC)
Based in part on observations with the NASA/ESA {\it Hubble Space Telescope},
obtained at the Space Telescope Science Institute, which is operated by the
Association of Universities for Research in Astronomy, In., under NASA
contract NAS 5-26555.

We thank Dr. R. Gredel, former director of Calar Alto, for supporting this
work and providing us with the telescope time for finishing it.

We thank support from the Spanish Plan Nacional de Astronom\'\i a program
AYA2005-09413-C02-02, of the Spanish Ministery of Education and Science and
the Plan Andaluz de Investigaci\'on of Junta de Andaluc\'{\i}a as research
group FQM322.

We thank Dr. R.Peletier, who kindly read the article cleaning the English.
His commments on the science content have helped us to increase the quality of
this article.


\appendix

\section{Individual Spectrum}
\label{app:1}

\begin{figure*}
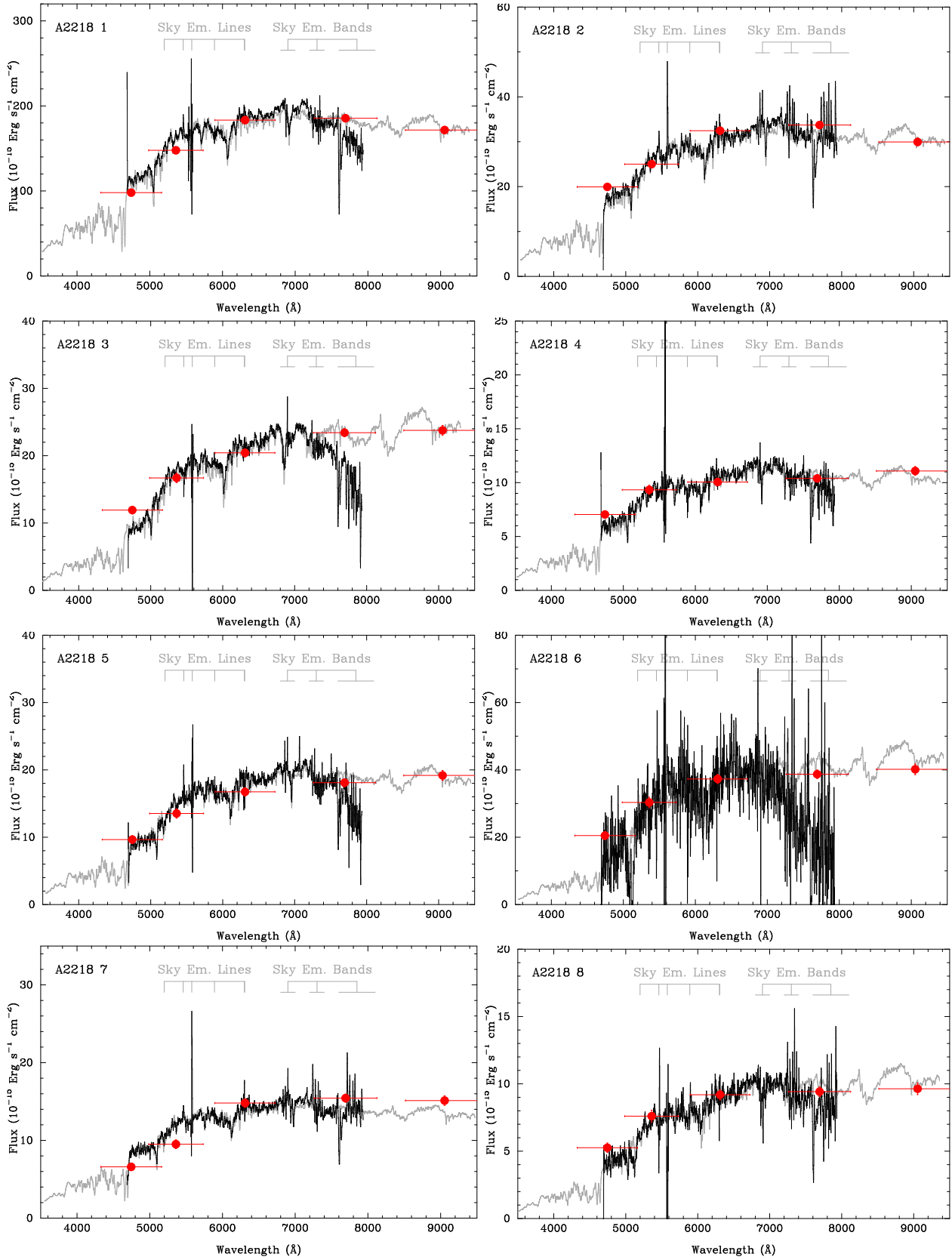

\includegraphics[width=5.5cm,angle=-90]{spec_A1.ps}
\includegraphics[width=5.5cm,angle=-90]{spec_A2.ps}
\includegraphics[width=5.5cm,angle=-90]{spec_A3.ps}
\includegraphics[width=5.5cm,angle=-90]{spec_A4.ps}
\includegraphics[width=5.5cm,angle=-90]{spec_A5.ps}
\includegraphics[width=5.5cm,angle=-90]{spec_A6.ps}
\includegraphics[width=5.5cm,angle=-90]{spec_A7.ps}
\includegraphics[width=5.5cm,angle=-90]{spec_A8.ps}
\caption{Individual extracted spectrum for each of the confirmed cluster
  members within the field-of-view of our IFS data (black solid line), each
  one labeled with the identification number listed in Table \ref{tab:red}.
  For each galaxy the spectrum is plot together with the multiband broad-band
  photometry derived from the HST/ACS F475W, F555W, F625W, F775W and
  F850LP-band images (red circles). The vertical error bars indicate the
  photometric errors, while the horizontal ones indicate the width of each
  band. he grey solid line shows the best fitted SSP model derived with the
  procedure described in Section \ref{sec:fit_SSP}.}
\label{fig:spec}       
\end{figure*}

Figure \ref{fig:spec} shows the indivual spectrum extracted from our IFS data
for each galaxies listed in the Table \ref{tab:red}.  For each galaxy the
spectrum is plot together with the multiband broad-band photometry derived
from the HST/ACS F475W, F555W, F625W, F775W and F850LP-band images (red
circles). The vertical error bars indicate the photometric errors, while the
horizontal ones indicate the width of each band. The wavelength of the
strongest night sky emission lines and bands are also indicated. At these
locations the unproper sky subtraction produce characteristic residuals that
can be seen in the spectra. In particular the wavelength range redder than
7200\AA\ is strongly affected, so we did not consider it in any of the
performed analysis. The grey solid line shows the best fitted SSP model
derived with the procedure described in Section \ref{sec:fit_SSP}.

\vfill

\vfill

\newpage

\newpage
\addtocounter{table}{-1}
\begin{figure*}
\includegraphics[width=5.5cm,angle=-90]{spec_A9.ps}
\includegraphics[width=5.5cm,angle=-90]{spec_A10.ps}
\includegraphics[width=5.5cm,angle=-90]{spec_A11.ps}
\includegraphics[width=5.5cm,angle=-90]{spec_A12.ps}
\includegraphics[width=5.5cm,angle=-90]{spec_A13.ps}
\includegraphics[width=5.5cm,angle=-90]{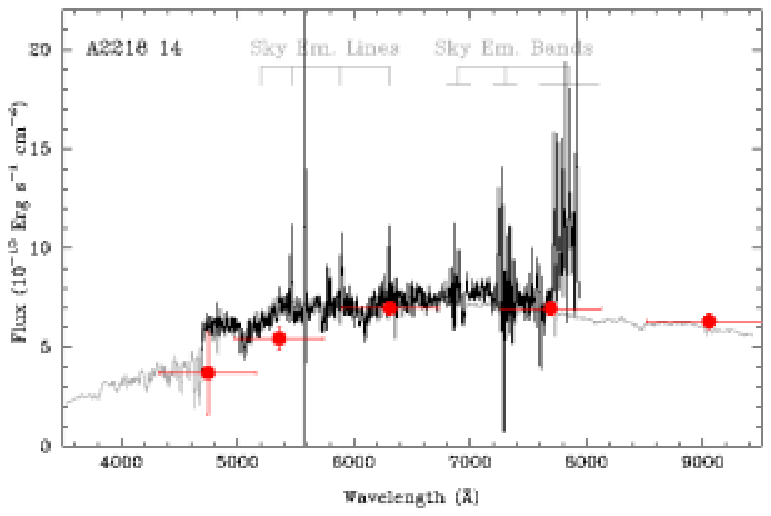}
\includegraphics[width=5.5cm,angle=-90]{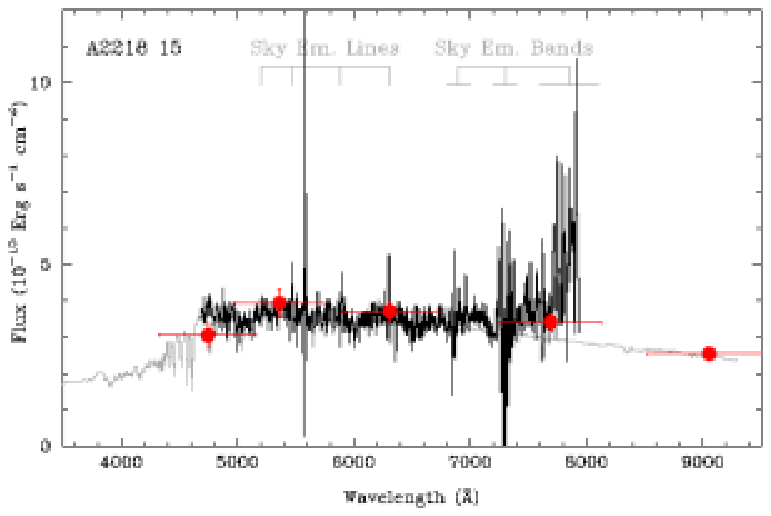}
\includegraphics[width=5.5cm,angle=-90]{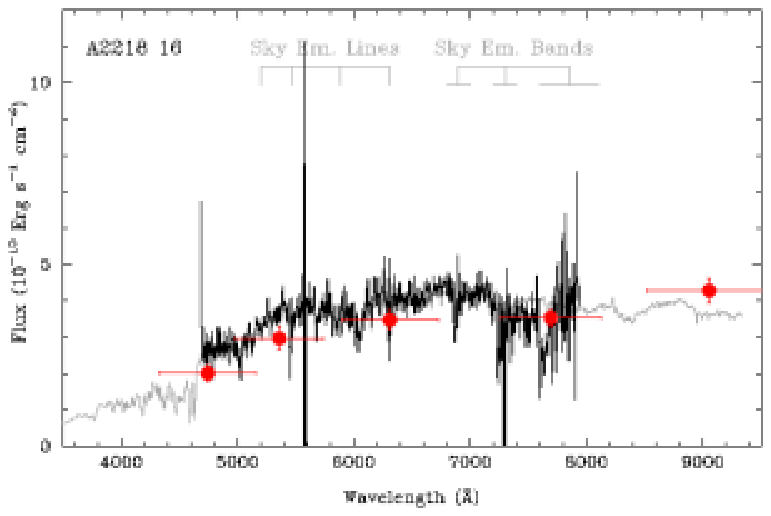}
\caption{Continue}
\end{figure*}
\newpage
\addtocounter{table}{-1}
\begin{figure*}
\includegraphics[width=5.5cm,angle=-90]{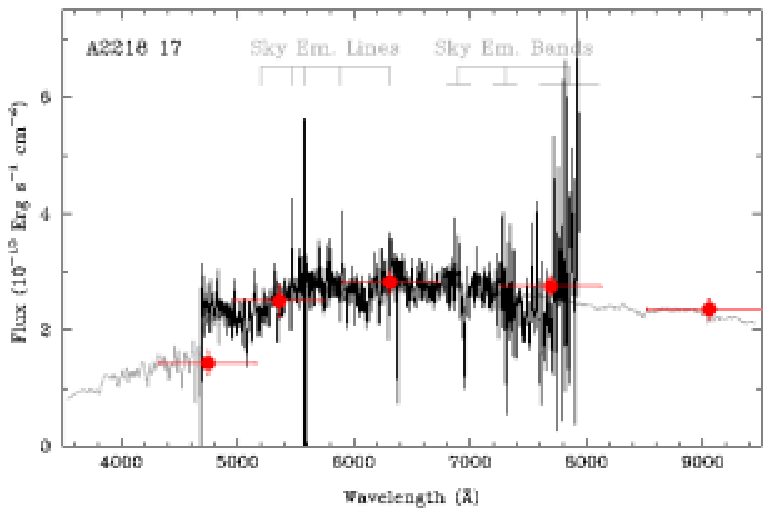}
\includegraphics[width=5.5cm,angle=-90]{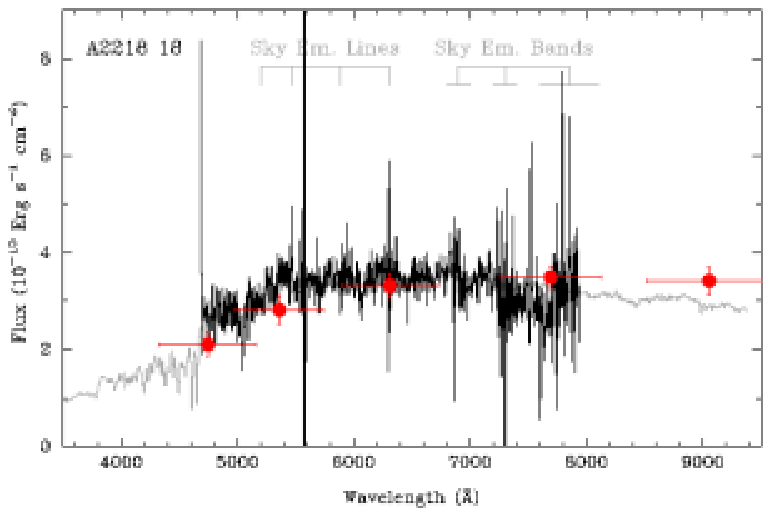}
\includegraphics[width=5.5cm,angle=-90]{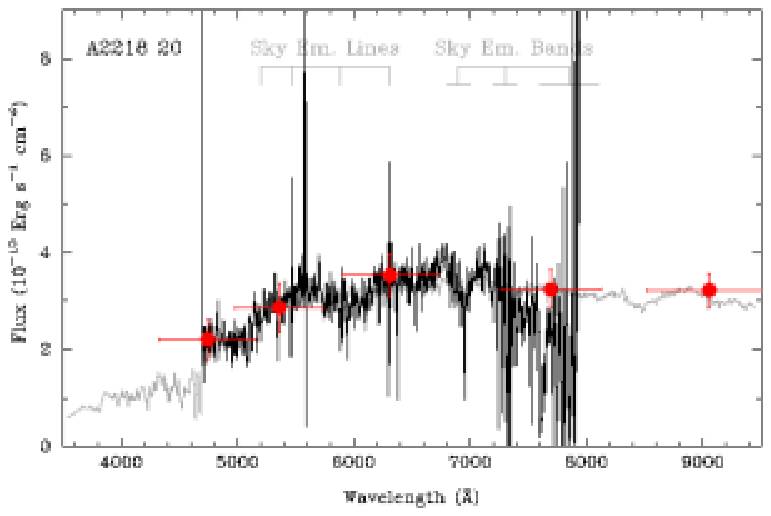}
\includegraphics[width=5.5cm,angle=-90]{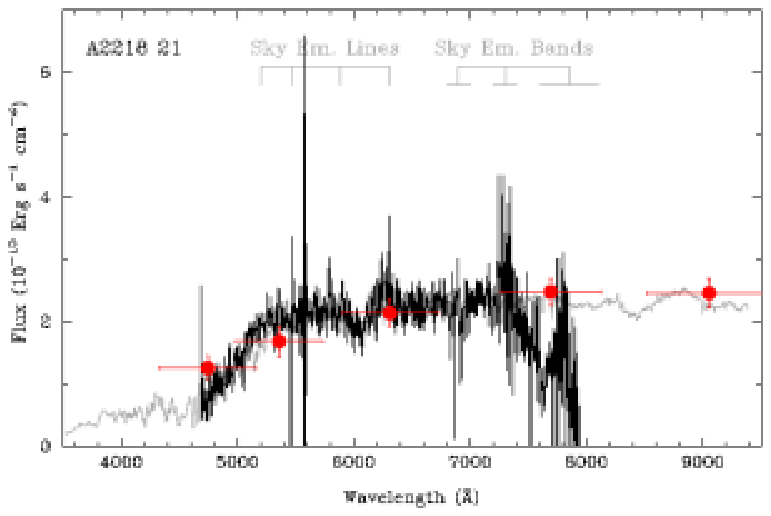}
\includegraphics[width=5.5cm,angle=-90]{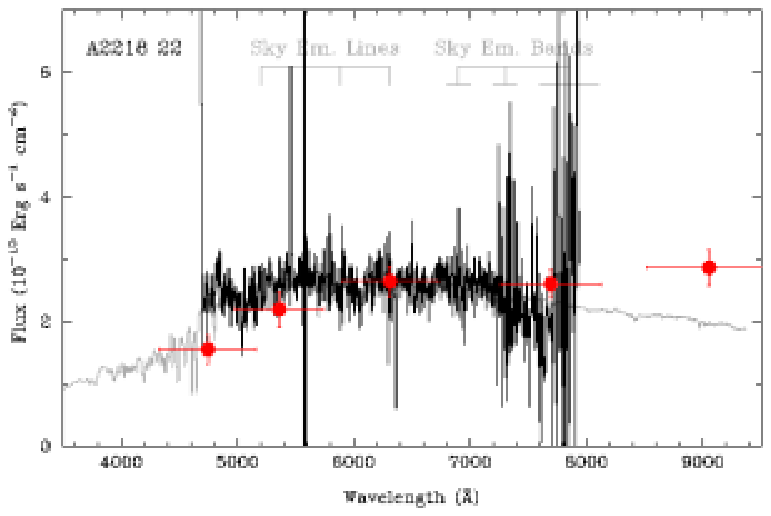}
\includegraphics[width=5.5cm,angle=-90]{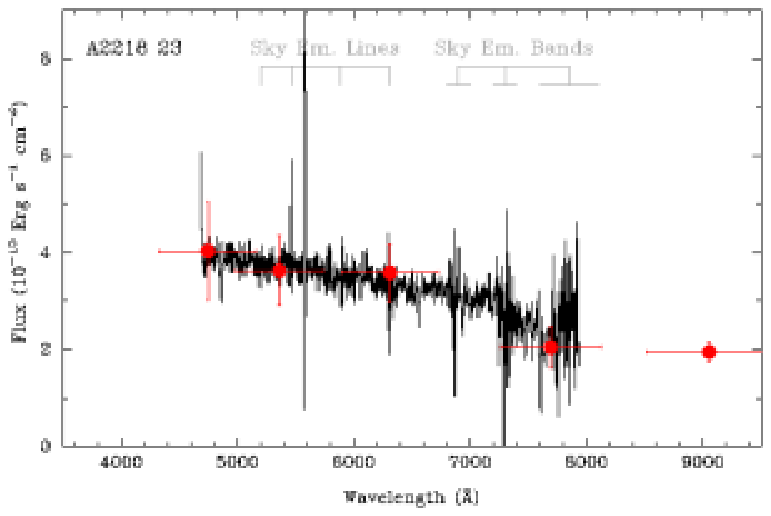}
\includegraphics[width=5.5cm,angle=-90]{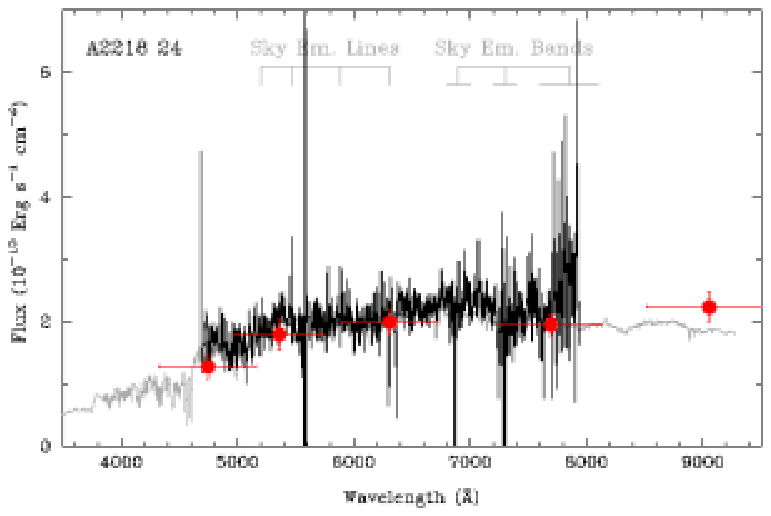}
\includegraphics[width=5.5cm,angle=-90]{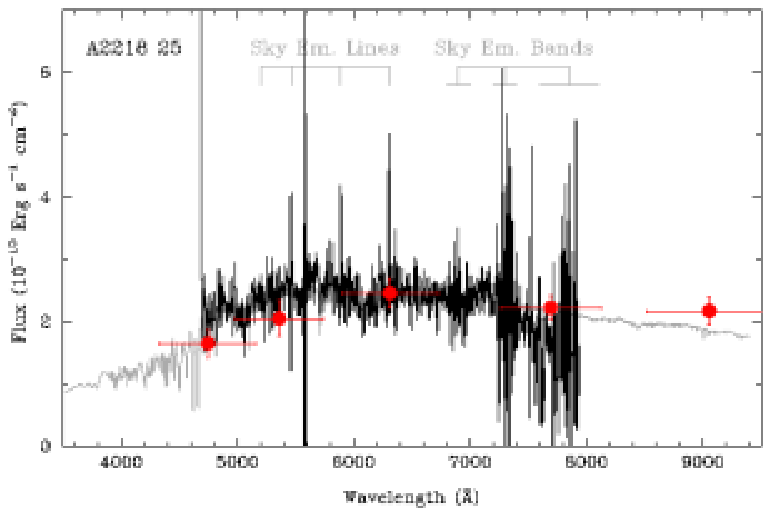}
\caption{Continue}
\end{figure*}
\newpage
\addtocounter{table}{-1}
\begin{figure*}
\includegraphics[width=5.5cm,angle=-90]{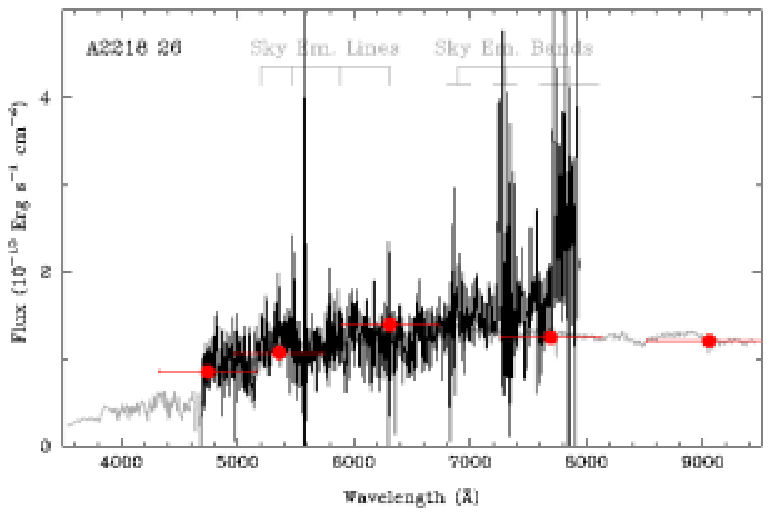}
\includegraphics[width=5.5cm,angle=-90]{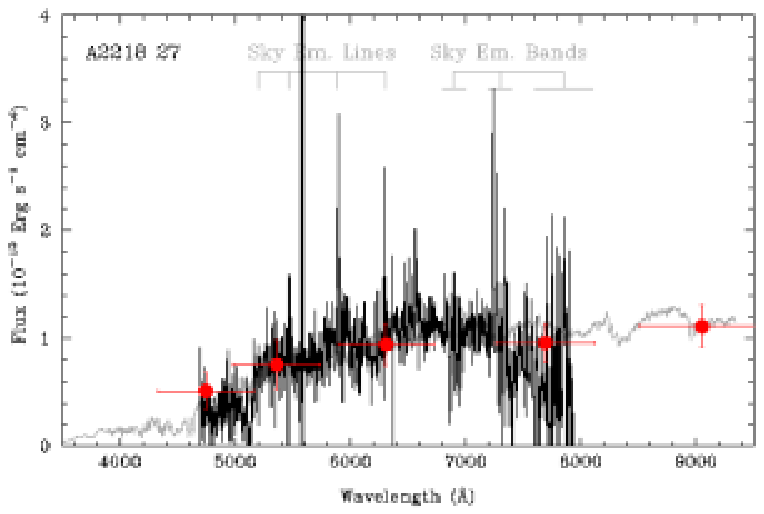}
\includegraphics[width=5.5cm,angle=-90]{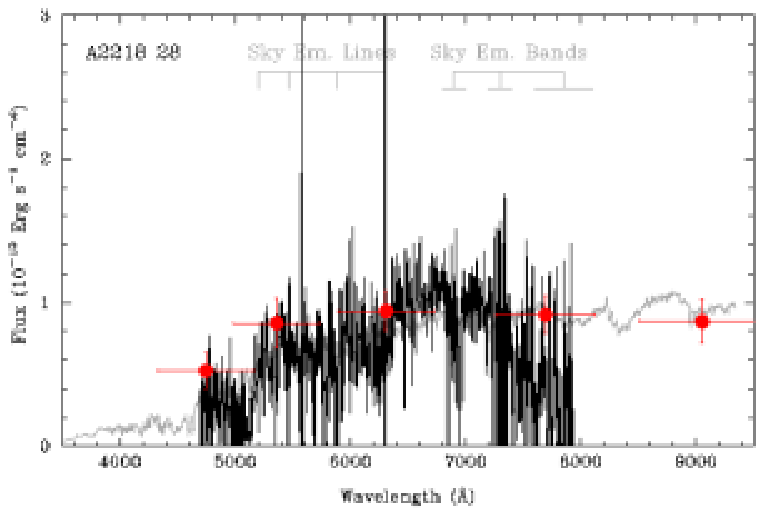}
\includegraphics[width=5.5cm,angle=-90]{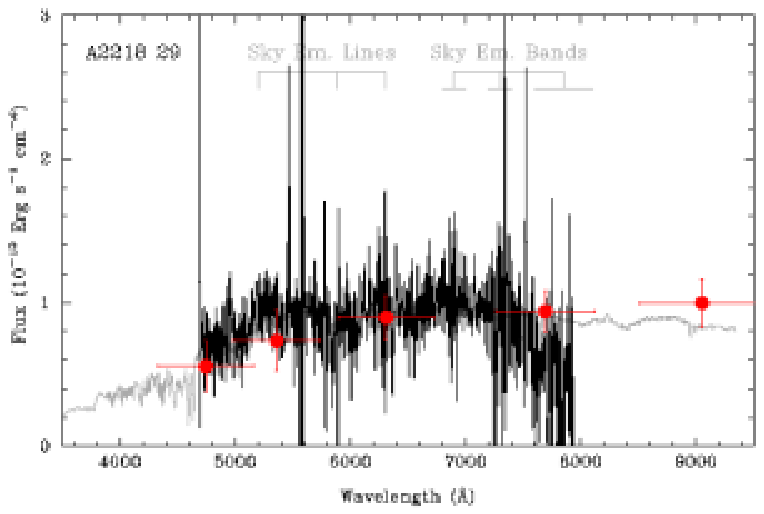}
\includegraphics[width=5.5cm,angle=-90]{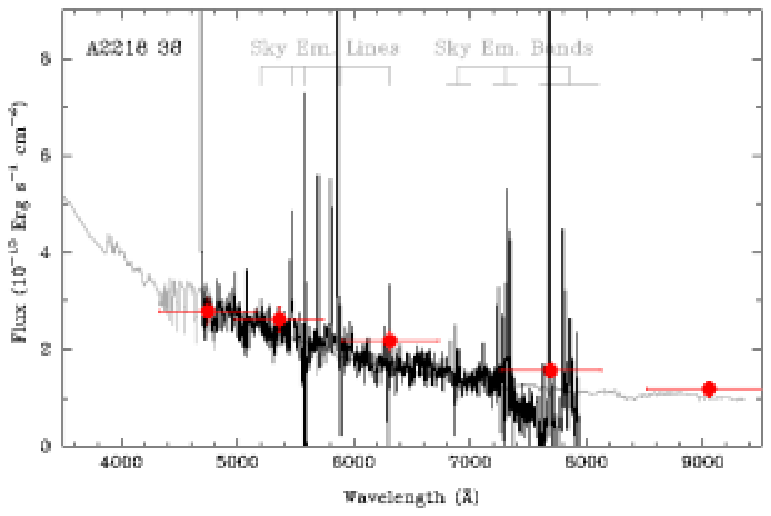}
\includegraphics[width=5.5cm,angle=-90]{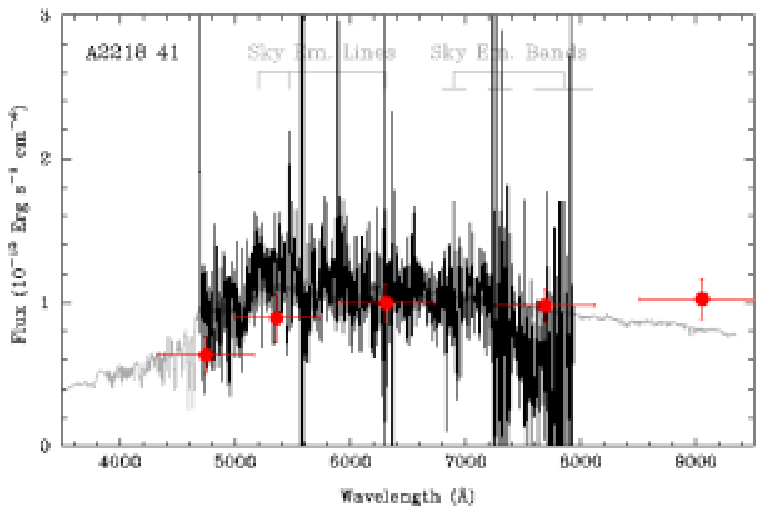}
\caption{Continue}
\end{figure*}

\vfill

\newpage

\section{Morphological Analysis and Spectra Extraction: Chromatic Effects}
\label{app:2}

The procedure to deblend and extract the spectra of the individual objects
from our IFS data relies on the assumption that it is possible to use the
morphological parameters derived by fitting galaxy profile models on the
F850LP-band image, using {\tt galfit}, to derive the photometry at any other
wavelength range, by forcing {\tt galfit} to recover the flux with the
morphological parameters fixed. To test this assumption use the HST/ACS
multiband image dataset, which cover all the wavelength range or our IFS data.

For each broad-band image of our dataset (F475W, F555W, F625W and F775W), we
performed the same morphological analysis that we already performed for the
F850LP-band, following the prescriptions described in Section \ref{sec:morph}.
In this first iteration we left {\tt galfit} to fit the morphological
parameters, as we did in the analysis of the F850LP-band. This procedure
produces a morphological catalogue for each of the considered images,
including the same parameters listed in Table \ref{tab:morph}.  In a second
iteration, the analysis was repeated but this time we used as input parameters
the ones derived by the analysis of the F850LP-band image. In this second step
the morphological parameters were fixed to those input values, and only the
photometry was fitted by {\tt galfit}. This procedure is basically the same
that we use to extract and deblend the spectra in our IFS data.

\begin{figure*}
\centering
\includegraphics[width=16cm,viewport=20 60 590 850,clip]{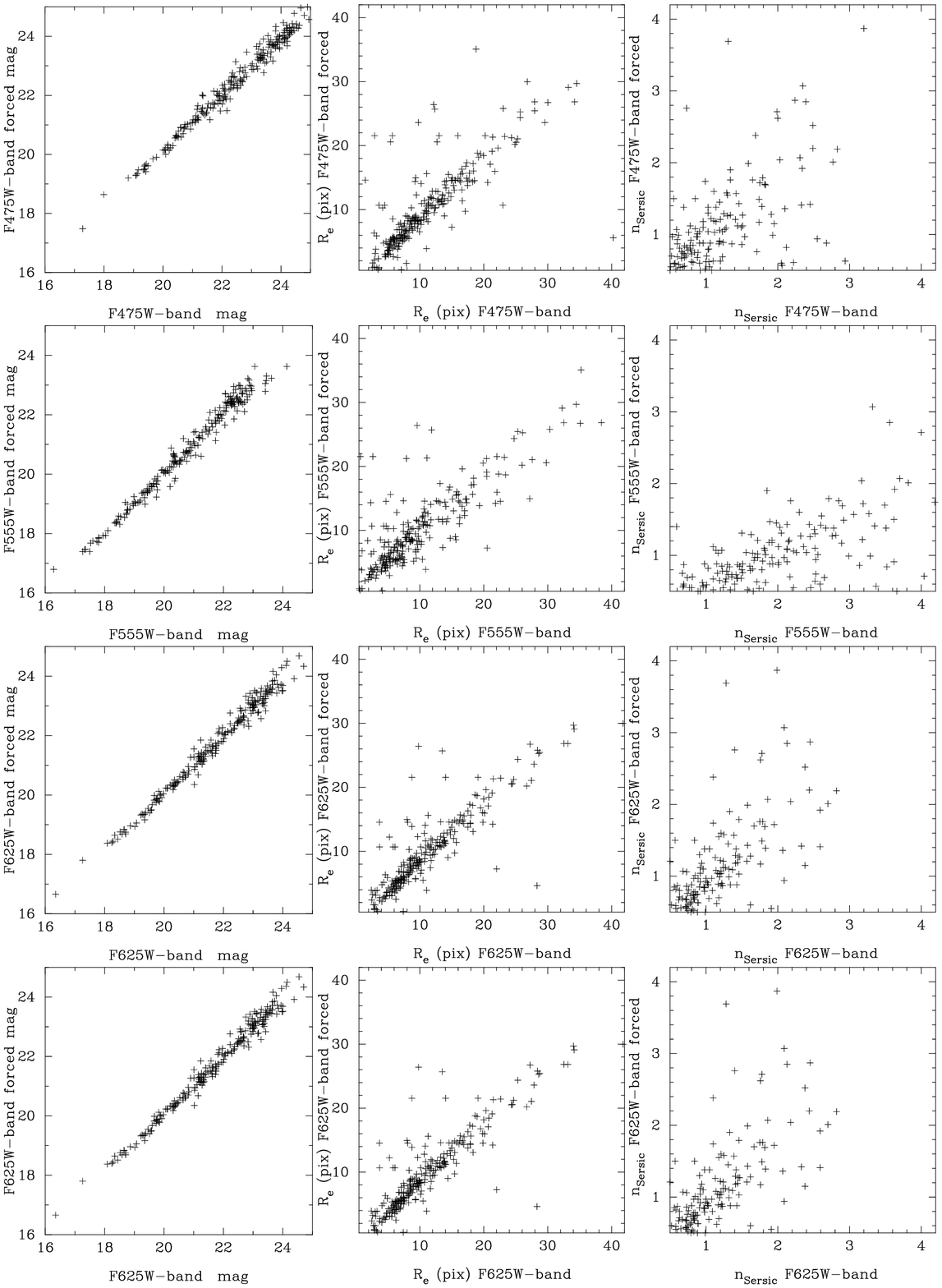}
\caption{{\it left-panesl:} F475W, F555W, F625W and F775W-band magnitudes derived by {\tt galfit}
  forcing the morphological parameters to be the ones derived from the
  morphological analysis of the F850LP-band image, as a function of the same
  magnitudes derived not imposing any condition to {\tt galfit} regarding the
  morphological parameters of the analyzed objects. {\it middle-panels:}
  similar comparison for the length-scales derived by {\tt galfit} for the
  F850LP-band image and for any other band .  {\it right-panels:} similar
  comparison for the S\'ersic indices.}
\label{fig:check}       
\end{figure*}

Similar results have been found for any band.
Figure \ref{fig:check} summarizes the results from the analysis. Left-panel
shows a comparison between the different bands magnitudes derived by {\tt
  galfit} forcing the morphological parameters to be the ones derived from the
morphological analysis of the F850LP-band image, and the magnitudes derived
not imposing any condition to {\tt galfit} regarding the morphological
parameters of the analyzed objects. The magnitudes derived using both methods
agree within $\pm$0.2 mags, without any clear trend or bias.  Middle-panel
shows the scale-length derived by {\tt galfit} for the F850LP-band image,
using in the forced fitting process over the different bands' images descrived
above, along the scale-length derived directly over this image by the
morphological analysis.  Although there are some outlayers, both scale-lengths
are similar in most of the cases. Right-panels shows a similar comparison
between the S\'ersic indices derived by the two fitting procedures.  In this
last case we found the biggest differences between both parameters, with a
considerable dispersion from the one-to-one relation. It is clear that there
is a chromatic effect over the morphology, as it is seen in the last panel,
however, this effect has a reduced impact on the derived scale-length, and
even less impact on the derived photometry. Thus, the basic assumption of our
procedure to deblend and extract the individual spectra of the targets from
our IFS data is valid.

\begin{figure*}
\centering
\includegraphics[width=16cm]{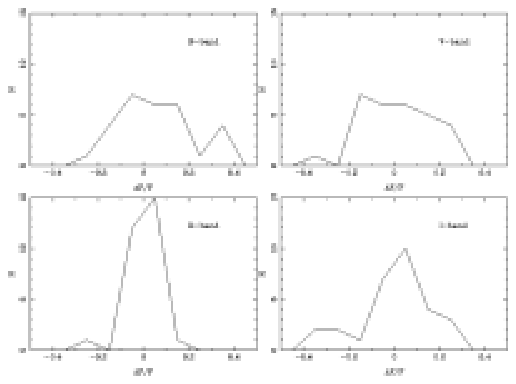}
\caption{Relative differences between the fluxes obtained for the $B$,$V$,$R$
  and $I$-bands derived from the IFS extracted spectra and those ones
 derived using the photometry obtained from the HST/ACS images.}
\label{fig:SYNTH}       
\end{figure*}

To perform a final test of the validity of our procedure and the accuracy of
our spectrophotometric data we synthetized the fluxes throughout the $B$, $V$,
$R$ and $I$-bands for each of our 28 cluster members with IFS data by
convolving their extracted spectra with the corresponding HST/ACS filter band
transmission curve, integrating their fluxes.  These fluxes are compared to
the corresponding ones derived using the HST/ACS based photometry listed in
Table \ref{tab:phot}. Figure \ref{fig:SYNTH} shows the histograms of the
relative differences between both fluxes for each band, $\Delta$F/F, that
basically corresponds to the error in the corresponding magnitude
($\sim$$e_{mag}$).  Due to the high accuracy of the HST/ACS photometry
(typical $e_{mag}\sim$0.05 mag for the considered objects) most of the
differences shown in the figure corresponds to spectrophotometric errors in
the IFS data. For all the bands both fluxes agree within a range of $\pm$0.3
mag, with a standard deviation of the difference of $\sim$0.15 mag.


\end{document}